\documentclass[12pt]{article}
\usepackage{makeidx}
\usepackage{amssymb}
\usepackage{amsfonts}
\usepackage{amsmath}
\usepackage{graphicx}
\usepackage{setspace}
\usepackage{authblk}
\usepackage{units}
\usepackage{appendix}
\usepackage[left=2cm,top=2.3cm,right=2cm,bottom=2.1cm]{geometry}
\usepackage{xcolor}
\usepackage[hyperfootnotes=false]{hyperref}

\renewcommand\familydefault\rmdefault

\hypersetup{colorlinks=true,linkcolor=black,citecolor=black,urlcolor=black}

\begin{document}

\title{{\textbf{Vacuum instability due to the creation of neutral Fermion with anomalous magnetic moment by magnetic-field inhomogeneities}}}
\author[1,2,3]{T. C. Adorno\thanks{\href{mailto:tg.adorno@gmail.com}{\texttt{tg.adorno@gmail.com}}\,,\,\href{mailto:adorno@hbu.edu.cn}{\texttt{adorno@hbu.edu.cn}}}}
\author[1,2]{Zi-Wang He\thanks{\href{mailto:1248049347@qq.com}{\texttt{1248049347@qq.com}}}}
\author[3,4]{S. P. Gavrilov\thanks{\href{mailto:gavrilovsergeyp@yahoo.com}{\texttt{gavrilovsergeyp@yahoo.com}}\,,\,\href{mailto:gavrilovsp@herzen.spb.ru}{\texttt{gavrilovsp@herzen.spb.ru}}}}
\author[3,5,6]{D. M. Gitman\thanks{\href{mailto:gitman@if.usp.br}{\texttt{gitman@if.usp.br}}}}
\affil[1]{\textit{Department of Physics, College of Physical Sciences and Technology, Hebei University, Wusidong Road 180, 071002, Baoding, China;}}
\affil[2] {\textit{Key Laboratory of High-precision Computation and Application of Quantum Field Theory of Hebei Province, Hebei University, Baoding, China;}}
\affil[3]{\textit{Department of Physics, Tomsk State University, Lenin Prospekt 36, 634050, Tomsk, Russia;}}
\affil[4]{\textit{Department of General and Experimental Physics, Herzen State
Pedagogical University of Russia, 191186, St. Petersburg, Russia;}}
\affil[5]{\textit{P. N. Lebedev Physical Institute, 53 Leninskiy prospekt,
119991, Moscow, Russia;}}
\affil[6]{\textit{Instituto de F\'{\i}sica, Universidade de S\~{a}o Paulo, Caixa Postal 66318, CEP 05508-090, S\~{a}o Paulo, S.P., Brazil.}}

\maketitle

\onehalfspacing

\begin{abstract}
We study neutral Fermions pair creation with anomalous magnetic moment from the vacuum by time-independent magnetic-field inhomogeneity as an external background. We show that the problem is technically reduced to the
problem of charged-particle creation by an electric step, for which the
nonperturbative formulation of strong-field QED is used. We consider a magnetic step given by an analytic function and whose inhomogeneity may vary from a ``gradual'' to a ``sharp'' field configuration. We obtain
corresponding exact solutions of the Dirac-Pauli equation with this field
and calculate pertinent quantities characterizing vacuum instability, such
as the differential mean number and flux density of pairs created from the
vacuum, vacuum fluxes of energy and magnetic moment. We show that the vacuum
flux in one direction is formed from fluxes of particles and antiparticles
of equal intensity and with the same magnetic moments parallel to the
external field. Backreaction to the vacuum fluxes leads to a smoothing of
the magnetic-field inhomogeneity. We also estimate critical magnetic field
intensities, near which the phenomenon could be observed.
\end{abstract}

\section{Introduction\label{Sec1}}

The violation of vacuum stability stimulated by external electromagnetic
fields is commonly associated with the possibility of such backgrounds
producing work on virtual pairs of particles and antiparticles. The most
well-known examples are electric-like fields, as they produce work on
charged particles and are able to tear apart electron/positron pairs from
the vacuum if the field amplitudes approach the so-called Schwinger critical
value $E_{\mathrm{c}}=m^{2}c^{3}/e\hslash \approx 1.3\times 10^{16}\,\mathrm{%
V}/\mathrm{cm}$ \cite{Schwinger51}. The phenomenon has been a subject of
intense investigation since the seminal works of Klein \cite{Klein29},
Sauter \cite{Sauter31a,Sauter31}, Heisenberg and Euler \cite{HeiEul36}, and
Schwinger \cite{Schwinger51}. An extensive discussion about the origin of
the effect, theoretical foundations, and experimental aspects can be found
in some reviews and monographs; see e.g. \cite%
{Greiner,Grib,FGS,Dunne04,ruffini,Dunne09,GelTan16,Piazza12,HegMouRaf14,AdoGavGit17}
and references therein.

Following the above interpretation, one may ask oneself about the
possibility that inhomogeneous macroscopic magnetic fields which, contrary
to homogeneous magnetic fields, produce a work on particles with a magnetic
moment, may create pairs from the vacuum. The answer to this question is
affirmative, provided the particles are neutral and have an anomalous magnetic
moment. Bearing in mind, first of all, very strong magnetic fields observed
in astrophysics, we can assume that this type of fields is practically
time-independent and steplike, that is, their gradient is always positive.
It must be said that the works available in the literature on calculating
such an effect, using sometimes inconsistent heuristic approaches, often
contradict each other \cite{Lin99,LeeYoo06,LeeYoo07,LeeYoo07b}. In this
regard, in work \cite{GavGit13}, it was shown that the problem could be
formally reduced to the calculation of charged particle creation from the
vacuum by stationary inhomogeneous electric fields of constant direction
(the so-called electric step potential gives this field). In particular, it
was shown that we are technically dealing with the calculation of an effect
similar to the well-known Klein effect, see Refs. \cite%
{Klein29,Sauter31a,Sauter31}. We recall that study of the effect began early
in the framework of relativistic quantum mechanics, see Ref. \cite{DomCal99}
for a review; its consistent nonperturbative description within QED was
given by Gavrilov and Gitman, in Refs. \cite{GavGit16,GavGit21}.

At present, there exist two types of particles enjoying the properties mentioned above:
the neutron and the neutrino. According to experimental data, neutrons have
a magnetic moment $\mu _{N}\approx -1.04187563(25)\times 10^{-3}\mu _{%
\mathrm{B}}$ \cite{NIST}, where $\mu _{\mathrm{B}}$ is the Bohr magneton. As
for neutrinos, there is not a general consensus because of the different
types of neutrinos, mechanism under which neutrinos acquires magnetic
moment, specific models, etc. Presently, experimental constraints range from 
$\mu _{\nu _{\tau }}<3.9\times 10^{-7}\mu _{\mathrm{B}}$ (for the tau
neutrino) \cite{donut01} until $\mu _{\nu _{e}}<2.9\times 10^{-11}\mu _{%
\mathrm{B}}$ (for the electron neutrino) \cite{gemma12}. Moreover, stringent
constraints obtained from astrophysical observations \cite%
{Raffelt90b,Raffelt90c,Raffelt92,Castel93,Catelan96,Viaux13} indicate that $%
\mu _{\nu }<\left( 2.6-4.5\right) \times 10^{-12}\mu _{\mathrm{B}}$ while
lower upper bounds, predicted by effective theories above the electroweak
scale, suggest that $\mu _{\nu }<10^{-14}\mu _{\mathrm{B}}$ \cite{Bell05}.
It is important to point out that for some theories beyond the Standard
Model (SM) \cite{Aboubrahim14}, it was reported that the magnetic moment for
the neutrinos lie within the range $\left( 10^{-12}-10^{-14}\right) \mu _{%
\mathrm{B}}$. For a more extensive discussion concerning experimental
aspects and theoretical predictions for neutrinos' electromagnetic
properties, see e.g. the reviews \cite%
{GiuStu09,Dvornikov11,BroGiuStu12,GiuStu15,GiuStu16} and references therein.

Vacuum instability effects due to inhomogeneous magnetic fields may be relevant to studies on dark matter. One of the reasons concerns the existence of sterile neutrinos, which may constitute dark matter and also couple to external magnetic fields through their electromagnetic properties. Presently, it is known that light sterile neutrinos appear in the low-energy effective theory in most extensions of the SM and, in principle, can have any mass, particularly in the range of $1\,\mathrm{eV}$. Sterile neutrinos with masses of several $%
\mathrm{keV}$ can account for cosmological dark matter, see e.g. Refs. \cite%
{Kuz09,Drewes17} for a recent review, and references therein. It is possible
that due to some new physics, the neutrino magnetic moment is big. Various
observational constraints on the magnetic moment $\mu $ of a dark matter
particle for masses $M$ in the range $1\,\mathrm{keV}$ to $100\,\mathrm{MeV%
}$ have been considered in Refs. \cite{Sig-etal04,Gard09}. The strongest
limits on $\mu$ emerge at the lightest mass scales. For example, if $%
M=m_{e}/10$, then $\left\vert \mu \right\vert <3.4\times 10^{-5}\mu _{B}$\
due to precision electroweak measurements.

Apart from neutral particles, it is important to mention that charged
particles can be created by inhomogeneous magnetic fields in de Sitter
spacetimes \cite{CruAnd16,AndCruPop18,AndCruPop20}. In Refs. \cite%
{CruAnd16,AndCruPop18} for instance, the authors calculated transition
amplitudes and probabilities of charged pair production by the magnetic
field of a magnetic dipole as an external background in de Sitter spacetime
perturbatively and shown that the corresponding probabilities are
nontrivial. In Ref. \cite{AndCruPop20}, the consideration was further
extended to the case where an external Coulomb field creates charged
particles. Theoretically, it is also known that pairs of particles and
antiparticles having a magnetic charge can be created by a constant and
homogeneous magnetic field \cite{AffMan82,AffAlvMan82}. This matter has
gained considerable attention lately due to the possibility of
monopole-antimonopole and axion-like particle pair production in the early
Universe \cite{Kobayashi21,LonVac15,Grifols99,Grifols02} and during
heavy-ion collisions \cite{GouHoRaj19,Rajantie19,GouHoRaj21}, especially
with the design of recent experiments aimed at detecting magnetic monopoles
in the Large Hadron Collider (LHC); e.g. \cite{Acharya16,Acharya18,Acharya19}
(see also Ref. \cite{MavMit20} for a review). In particular, the mechanism
of monopole-antimonopole pair production provides an explanation for the
dissipation of primordial magnetic fields consistent with gamma ray
observations of intergalactic magnetic fields \cite%
{Tavecchio10,NerVov10,Dermer11}. It has been discussed that even heavy
monopoles (with masses $\sim 10^{16}\,\mathrm{GeV}$) could be pair produced
in the early Universe \cite{Kobayashi21}. Besides the aforementioned types
of particles and external backgrounds, it should be noted that
neutrino/antineutrino ($\nu \bar{\nu}$) pairs may also be created from the
vacuum by dense matters as external backgrounds. For example, it was
reported some years ago that $\nu \bar{\nu}$ pairs can be created in Neutron
stars \cite{Loeb90,Kalch98} due to the interaction with the background
matter. In these works, the authors considered time-independent matter
densities and calculated $\nu \bar{\nu}$ pair production rate in analogy
with Schwinger's result for electron-positron pair production by a constant
electric field \cite{Schwinger51}. It was also reported that $\nu \bar{\nu}$
pairs can also be created due to a coherent interaction with a dense medium 
\cite{KieWei97}. Some years later, $\nu \bar{\nu}$ pair production rates for
time-dependent matter densities were calculated nonperturbatively through
semiclassical methods in Ref. \cite{KusPos02} and perturbatively in Ref. 
\cite{Koers05}, using the $S$-matrix formalism. More recently, some of us 
\cite{DvoGavGit14} considered a consistent nonperturbative formulation for
calculating $\nu \bar{\nu}$ pair creation from the vacuum by a
time-dependent background matter. It was also demonstrated in Ref. \cite%
{Dvornikov15} that $\nu \bar{\nu}$ pairs could be created from the vacuum by
an accelerated matter due to the neutrino electroweak interaction with
background Fermions.

Understanding the mechanism responsible for neutral Fermion pair production
by intense magnetic fields may be particularly important to comprehend the
physics of highly magnetized astrophysical objects and events occurring in
the Universe. For example, it was reported a few years ago \cite%
{Lai01,Akiyama03,Mereghetti08}, that magnetic fields of the order of $%
10^{16}-10^{18}\,\mathrm{G}$ could be generated during a supernova explosion
or in the vicinity of magnetars. Moreover, ultra-intense magnetic fields (of
order up to $10^{20}\ \mathrm{G}$) can be produced at the core of compact
magnetars \cite{Ferrer10-11}. Based on experimental values for the neutron
mass and its magnetic moment \cite{NIST}, it is possible that such objects
can produce pairs of neutrons/antineutrons, and it may affect their inner
dynamics. Another important aspect of the present study concerns the
possibility of inspecting conditions for neutrino pair production based on
predictions for the neutrinos magnetic moment\footnote{%
loop-induced magnetic moment.} supplied by the SM $\mu _{\nu }\approx
3.2\times 10^{-19}\mu _{\mathrm{B}}\times \left( m_{\nu }/1\,\mathrm{eV}%
\right) $ \cite{GiuStu09,GiuStu15}. Considering an acceptable range of values to neutrinos masses,
say $1\,\mathrm{eV}-10^{-2}\,\mathrm{eV}$, we discuss below that neutrinos with such small magnetic moments can be created only by inhomogeneous magnetic fields several orders above the astronomical
scale. The situation is very different for neutrinos with larger magnetic moments, which could be already created from the vacuum by magnetic fields of order $10^{16}\,\mathrm{G}$. Thus, neutrino pair production by intense
magnetic-field inhomogeneity may indirectly indicate that the SM
must be extended in order to account for larger magnetic moments, possibly
near the experimental reach.

The mechanism discussed in this work corresponds to the most fundamental
process driven by the interaction between neutral Fermions with
electromagnetic backgrounds, namely the creation of neutral Fermion pairs
with anomalous magnetic moment from the vacuum by an inhomogeneous external
magnetic field. The process in consideration is particularly different from
neutrino pair production by an electron in a constant magnetic field, in
which neutrino/antineutrino pairs are created as a decay process of the
initial electron stimulated by the magnetic field; see e.g. Ref. \cite%
{GiuStu15} and references therein. It was demonstrated in \cite{GavGit13}
that a quantization in terms of neutral particles and antiparticles is
possible in terms of the states with well-defined spin polarization. In this
case, the problem can be technically reduced to the problem of
charged-particle creation by electric fields given by step potentials (in
short, electric steps) for which a nonperturbative formulation in QED \cite{GavGit16,GavGit21} can be used. This formulation is based on the possibility
of finding exact solutions to the Dirac-Pauli equation with steplike
magnetic fields. As an example, neutral Fermion creation from the vacuum by
a linearly growing magnetic field was considered in Ref. \cite{GavGit13}. In
the present article, we develop the technique proposed in Ref. \cite%
{GavGit13} taking into account recent theoretical constructions \cite{GavGit16,GavGit21}. Based on these developments, we study neutral Fermion pair production from the vacuum by the magnetic step given by an analytic function that enables studying the role of the field inhomogeneity on pair production. In Sec. \ref{Sec2} we
demonstrate how to reduce the problem under consideration to the problem of
charged-particle creation by an electric step. In Sec. \ref{Sec2b},
we describe the field in consideration and construct corresponding 
\textrm{in}- and \textrm{out}-solutions of the Dirac-Pauli equation with
this field. With the aid of these solutions, we find differential and total
quantities characterizing vacuum instability. In Secs. \ref{Sec3} and \ref{Sec3.1}, we present these quantities and scrutinize their behavior when the field varies ``gradually" or ``sharply" along the inhomogeneity direction. Numerical estimates to the critical field are given. In Sec. \ref{Sec4}, we find vacuum fluxes of energy and
the magnetic moment produced by the magnetic-field inhomogeneity. Sec. \ref{Sec5}
is devoted to the concluding remarks. In this work we consider the
four-dimensional Minkowski spacetime, parameterized by coordinates $X=\left(
X^{\mu }\,,\ \mu =0,i\right) =\left( t,\mathbf{r}\right) $, $t=X^{0}$, $%
\mathbf{r}=X^{i}=\left( x,y,z\right) $, $i=1,2,3$, and metric tensor $\eta
_{\mu \nu }=\mathrm{diag}\left( +1,-1,-1,-1\right) $. We also employ natural
units, in which $\hslash =1=c$.

\section{Solutions of the Dirac-Pauli equation with well-defined spin
polarization\label{Sec2}}

In this section we present general considerations on the Dirac-Pauli
equation with inhomogeneous magnetic fields. In particular, we discuss the
spinor structure of solutions and their asymptotic properties at specific
remote distances. Such properties are important to correctly classify
solutions and to define quantities characterizing vacuum instability, as
discussed in Sec. \ref{Sec3}.

The motion of a relativistic spin $1/2$ neutral particle with anomalous
magnetic moment $\mu $, mass $m$, in external electromagnetic fields is
described by the relativistic wave equation%
\begin{equation}
\left( i\gamma ^{\mu }\partial _{\mu }-m-\frac{1}{2}\mu \sigma ^{\mu \nu
}F_{\mu \nu }\right) \psi \left( X\right) =0\,,  \label{1}
\end{equation}%
which is conventionally called the Dirac-Pauli (DP) equation\footnote{%
See the textbook \cite{BagGit14} for a more detailed consideration of the
Dirac-Pauli equation and its solutions to a wider class of electromagnetic
backgrounds.} \cite{Pauli41}. Here $\psi \left( X\right) $ is a four spinor, 
$\gamma ^{\mu }=\left( \gamma ^{0},\boldsymbol{\gamma }\right) $ are Dirac
matrices, $\sigma ^{\mu \nu }=\left( i/2\right) \left[ \gamma ^{\mu },\gamma
^{\nu }\right] _{-}$, $F_{\mu \nu }=\partial _{\mu }A_{\nu }-\partial _{\nu
}A_{\mu }$ is the electromagnetic field strength tensor, and $\mu $\ should
be understood as the algebraic value of the magnetic moment (e.g., $\mu
=-\left\vert \mu _{N}\right\vert $\ for a neutron). In what follows we
consider external fields of a specific type, corresponding to a
time-independent magnetic field oriented along the positive direction of the 
$z$-axis, inhomogeneous along the $y$-direction, $\mathbf{B}\left( \mathbf{r}%
\right) =\left( 0,0,B_{z}\left( y\right) \right) $, and homogeneous at
remote distances, $B_{z}\left( \pm \infty \right) =\mathrm{const}$.
Moreover, it is assumed that its gradient is always positive $\partial
_{y}B\left( y\right) \geq 0$,$\ \forall \,y\in \left( -\infty ,+\infty
\right) $, meaning that $B_{z}\left( +\infty \right) >B_{z}\left( -\infty
\right) $ and that the field is genuinely a step. We conveniently refer to
fields of this type as steplike magnetic fields or simply as magnetic steps.

The general method for solving the DP equation (\ref{1}) with steplike
magnetic fields was presented before by two of us in Ref. \cite{GavGit13}.
We recall some properties of the DP equation with such fields and present new details that simplify the spinor structure
of the solutions. In the Schr\"{o}dinger form, the DP equation (\ref{1})
reads%
\begin{equation}
i\partial _{t}\psi \left( X\right) =\hat{H}\psi \left( X\right) \,,\ \ \hat{H%
}=\gamma ^{0}\left( \gamma ^{3}\hat{p}_{z}+\Sigma _{z}\hat{\Pi}_{z}\right)
\,.  \label{2.1}
\end{equation}%
Here $\Sigma _{z}=i\gamma ^{1}\gamma ^{2}$ and%
\begin{equation}
\hat{\Pi}_{z}=\Sigma _{z}\left( \boldsymbol{\gamma }\mathbf{\hat{p}}_{\perp
}+m\right) -\mathbb{I}\mu B_{z}\left( y\right) \,,  \label{2.2}
\end{equation}%
is an integral of motion spin-operator, $\left[ \hat{\Pi}_{z},\hat{H}\right]
_{-}=0$. The subscript \textquotedblleft $\perp $\textquotedblright\ labels
quantities perpendicular to the field, e.g. $\mathbf{\hat{p}}_{\perp
}=\left( \hat{p}_{x},\hat{p}_{y}\right) $, and $\mathbb{I}$ denotes the $%
4\times 4$ identity matrix. Since the operators $\hat{p}_{0}$, $\hat{p}_{x}$%
, and $\hat{p}_{z}$\ are compatible with the Hamiltonian (and also with $%
\hat{\Pi}_{z}$), the DP spinor admits the general form $\psi _{n}\left(
X\right) =\exp \left( -ip_{0}t+ip_{x}x+ip_{z}z\right) \psi _{n}\left(
y\right) $, where $\psi _{n}\left( y\right) $ depends exclusively on $y$ and
obeys the eigenvalue equation:%
\begin{eqnarray}
&&\hat{\Pi}_{z}\psi _{n}\left( X\right) =e^{-ip_{0}t+ip_{x}x+ip_{z}z}\Pi
_{z}\psi _{n}\left( y\right) \,,\ \ \Pi _{z}\psi _{n}\left( y\right)
=s\omega \psi _{n}\left( y\right) \,,\ \ s=\pm 1\,,  \notag \\
&&\Pi _{z}=\hat{\pi}_{z}-\mathbb{I}\mu B_{z}\left( y\right) \,,\ \ \hat{\pi}%
_{z}=\Sigma _{z}\left( \gamma ^{1}p_{x}+\gamma ^{2}\hat{p}_{y}+m\right) \,,
\label{2.4}
\end{eqnarray}%
By acting the squared Hamiltonian operator (\ref{2.1}) onto $\psi _{n}\left(
X\right) $, we observe that the total particle's energy $p_{0}$,
longitudinal momentum $p_{z}$, and $\omega $ are interrelated, $%
p_{0}^{2}=\omega ^{2}+p_{z}^{2}\rightarrow p_{0}=\omega \sqrt{%
1+p_{z}^{2}/\omega ^{2}}$, $\mathrm{sgn}\left( p_{0}\right) =\mathrm{sgn}%
\left( \omega \right) $. This relation indicates that $\omega $ is the
transverse\footnote{%
that is, the total energy on the $xy$ plane.} part of the total energy.
Thanks to this identity, we may introduce an additional operator%
\begin{equation}
\hat{R}=\hat{H}\hat{\Pi}_{z}^{-1}\left[ \mathbb{I}+\left( \hat{p}_{z}\hat{\Pi%
}_{z}^{-1}\right) ^{2}\right] ^{-1/2}\,,  \label{2.6}
\end{equation}%
which is also an integral of motion and commutes with all previous
operators. In particular, this operator implies that $\psi _{n}\left(
y\right) $ also obeys the eigenvalue equation%
\begin{eqnarray}
&&\hat{R}\psi _{n}\left( X\right) =e^{-ip_{0}t+ip_{x}x+ip_{z}z}R\psi
_{n}\left( y\right) \,,\ \ R\psi _{n}\left( y\right) =s\psi _{n}\left(
y\right) \,,  \notag \\
&&R=\Upsilon \gamma ^{0}\left( \Sigma _{z}+\frac{sp_{z}}{\omega }\gamma
^{3}\right) \,,\ \  \Upsilon =\frac{1}{\sqrt{1+p_{z}^{2}/\omega ^{2}}}\,.
\label{2.8}
\end{eqnarray}%
As a result, we select $\hat{p}_{x}$, $\hat{p}_{z}$, $\hat{\Pi}_{z}$, $\hat{R%
}$ as the complete set of commuting operators, whose eigenvalues are $%
n=\left( p_{x},p_{z},\omega ,s\right) $.

The set of equations (\ref{2.4}) and (\ref{2.8}) are simultaneously
satisfied choosing $\psi _{n}\left( y\right) $ in the form%
\begin{equation}
\psi _{n}\left( y\right) =\left( \mathbb{I}+sR\right) \left[ \hat{\pi}_{z}+%
\mathbb{I}\left( \mu B_{z}\left( y\right) +s\omega \right) \right] \varphi
_{n,\chi }\left( y\right) \upsilon _{\kappa }^{\left( \chi \right) }\,,
\label{2.9}
\end{equation}%
where $\varphi _{n,\chi }\left( y\right) $ are functions while $\upsilon
_{\kappa }^{\left( \chi \right) }$ belongs to a set of four constant
spinors, satisfying the eigenvalue equations%
\begin{equation}
i\gamma ^{1}\upsilon _{\kappa }^{\left( \chi \right) }=\chi \upsilon
_{\kappa }^{\left( \chi \right) }\,,\ \ \gamma ^{0}\gamma ^{2}\upsilon
_{\kappa }^{\left( \chi \right) }=\kappa \upsilon _{\kappa }^{\left( \chi
\right) }\,,\ \ \chi =\pm 1\,,\ \ \kappa =\pm 1\,,  \label{2.10}
\end{equation}%
and the orthonormality conditions $\upsilon _{\kappa ^{\prime }}^{\left(
\chi ^{\prime }\right) \dagger }\upsilon _{\kappa }^{\left( \chi \right)
}=\delta _{\chi ^{\prime }\chi }\delta _{\kappa ^{\prime }\kappa }$.
Substituting the spinor (\ref{2.9}) into (\ref{2.4}) one finds that the
scalar functions $\varphi _{n,\chi }\left( y\right) $ are solutions of the
second-order ordinary differential equation%
\begin{equation}
\left\{ -\frac{d^{2}}{dy^{2}}-\left[ s\omega +\mu B_{z}\left( y\right) %
\right] ^{2}+\pi _{x}^{2}+i\mu \chi B_{z}^{\prime }\left( y\right) \right\}
\varphi _{n,\chi }\left( y\right) =0\,,\ \ \pi _{x}^{2}=m^{2}+p_{x}^{2}\,.
\label{2.11}
\end{equation}%
Although the separation of degrees-of-freedom given by Eq. (\ref{2.9})
simplifies the structure of the solutions, it is important to point out that
none of the operators listed in (\ref{2.10}) are integrals of motion. Thus,
it is possible to select any spinor of the basis $\upsilon _{\kappa
}^{\left( \chi \right) }$ to study pair creation. To keep expressions as
general as possible, we leave this choice arbitrary in all calculations
below.

At this point, it is worth discussing some general features of neutral
Fermions in the external field and properties of the solutions. Because of
the spectrum of the spin operator $\hat{R}$, there are two species of
neutral Fermions differing by the value of $s$--one species for which $s=+1$
and another for which $s=-1$. Consequently, the potential energy of a
neutral Fermion in this field is given by $U_{s}\left( y\right) =sU\left(
y\right) $, where $U\left( y\right) =-\mu B_{z}\left( y\right) $.\textrm{\ }%
To facilitate subsequent discussions, it is convenient to select a fixed
sign for particle magnetic moment. Thus, from now on, we choose a Fermion
with a negative magnetic moment as the main particle, $\mu =-\left\vert \mu
\right\vert $. Since the external field increases monotonically with $y$
(its gradient is positive, as stated before), the maximum potential\textrm{\ 
}energy that may be experienced by the Fermion is determined by the
magnitude of the step%
\begin{equation}
\mathbb{U}\equiv U_{\mathrm{R}}-U_{\mathrm{L}}>0\,,  \label{2.12}
\end{equation}%
which is essentially the difference between the asymptotic values $U_{%
\mathrm{R}}=U\left( +\infty \right) $,\ $U_{\mathrm{L}}=U\left( -\infty
\right) $\ and is positive, by definition\footnote{%
The labels \textquotedblleft \textrm{L}\textquotedblright\ and
\textquotedblleft \textrm{R}\textquotedblright\ mean \textquotedblleft
asymptotic left region $y\rightarrow -\infty $\textquotedblright\ and
\textquotedblleft asymptotic right region $y\rightarrow +\infty $%
\textquotedblright , respectively.}. Thus, for Fermions with $s=+1$, the
magnitude of the potential\ step\ is $\mathbb{U}=U_{+1}\left( \mathrm{R}%
\right) -U_{+1}\left( \mathrm{L}\right) $ while for Fermions with $s=-1$ it
is $\mathbb{U}=U_{-1}\left( \mathrm{L}\right) -U_{-1}\left( \mathrm{R}%
\right) $, where $U_{s}\left( \mathrm{L/R}\right) =sU_{\mathrm{L/R}}$. At
remote distances--where the field can be considered as\emph{\ }homogeneous
and no longer accelerates particles--the term proportional to $\chi $ in Eq.
(\ref{2.11}) is absent. Therefore, solutions of Eq. (\ref{2.11})
intrinsically have well-defined \textit{left}\ $\ _{\zeta }\varphi _{n,\chi
}\left( y\right) $ and \textit{right}\ $\ ^{\zeta }\varphi _{n,\chi }\left(
y\right) $\ asymptotic forms%
\begin{eqnarray}
\ _{\zeta }\varphi _{n,\chi }\left( y\right) &=&\ _{\zeta }\mathcal{N}\exp
\left( i\zeta \left\vert p^{\mathrm{L}}\right\vert y\right) \,,\ \ \zeta =%
\mathrm{sgn}\left( p^{\mathrm{L}}\right) \,,\ \ y\rightarrow -\infty \,, 
\notag \\
\ ^{\zeta }\varphi _{n,\chi }\left( y\right) &=&\ ^{\zeta }\mathcal{N}\exp
\left( i\zeta \left\vert p^{\mathrm{R}}\right\vert y\right) \,,\ \ \zeta =%
\mathrm{sgn}\left( p^{\mathrm{R}}\right) \,,\ \ y\rightarrow +\infty \,,
\label{2.14}
\end{eqnarray}%
in which $\ _{\zeta }\mathcal{N}$, $\ ^{\zeta }\mathcal{N}$ are
normalization constants,\ $\left\vert p^{\mathrm{L/R}}\right\vert $ are $y$%
-components of Fermions momenta at corresponding remote regions,%
\begin{equation}
\left\vert p^{\mathrm{L/R}}\right\vert =\sqrt{\left[ s\pi _{s}\left( \mathrm{%
L/R}\right) \right] ^{2}-\pi _{x}^{2}}\,,\ \ \pi _{s}\left( \mathrm{L/R}%
\right) =\omega -sU_{\mathrm{L/R}}\,,  \label{2.15}
\end{equation}%
and $\pi _{s}\left( \mathrm{L/R}\right) $ are their transverse kinetic
energies at remote areas. Correspondingly, we may introduce the
asymptotically-\textit{left} $\ _{\zeta }\psi _{n}\left( X\right) =\exp
\left( -ip_{0}t+ip_{x}x+ip_{z}z\right) \ _{\zeta }\psi _{n}\left( y\right) $
and the asymptotically-\textit{right} $^{\zeta }\psi _{n}\left( X\right)
=\exp \left( -ip_{0}t+ip_{x}x+ip_{z}z\right) \ ^{\zeta }\psi _{n}\left(
y\right) $\ sets of DP spinors%
\begin{eqnarray}
\ _{\zeta }\psi _{n}\left( y\right) &=&\ _{\zeta }\mathcal{N}e^{i\zeta
\left\vert p^{\mathrm{L}}\right\vert y}\left( \mathbb{I}+sR\right) \left[
\Sigma _{z}\left( \gamma ^{1}p_{x}+m\right) +\mathbb{I}\left( s\pi
_{s}\left( \mathrm{L}\right) -\chi \zeta \left\vert p^{\mathrm{L}%
}\right\vert \right) \right] \upsilon _{\kappa }^{\left( \chi \right) }\,,\
\ y\rightarrow -\infty \,,  \notag \\
\ ^{\zeta }\psi _{n}\left( y\right) &=&\ ^{\zeta }\mathcal{N}e^{i\zeta
\left\vert p^{\mathrm{R}}\right\vert y}\left( \mathbb{I}+sR\right) \left[
\Sigma _{z}\left( \gamma ^{1}p_{x}+m\right) +\mathbb{I}\left( s\pi
_{s}\left( \mathrm{R}\right) -\chi \zeta \left\vert p^{\mathrm{R}%
}\right\vert \right) \right] \upsilon _{\kappa }^{\left( \chi \right) }\,,\
\ y\rightarrow +\infty \,,  \label{2.16}
\end{eqnarray}%
which, in turn, obey the eigenvalue equations%
\begin{eqnarray}
&&\hat{p}_{y}\ _{\zeta }\psi _{n}\left( X\right) =\zeta \left\vert p^{%
\mathrm{L}}\right\vert \ _{\zeta }\psi _{n}\left( X\right) \,,\ \ \hat{h}%
_{\perp }^{\mathrm{kin}}\ _{\zeta }\psi _{n}\left( X\right) =s\pi _{s}\left( 
\mathrm{L}\right) \ _{\zeta }\psi _{n}\left( X\right) \,,\ \ y\rightarrow
-\infty \,,  \notag \\
&&\hat{p}_{y}\ ^{\zeta }\psi _{n}\left( X\right) =\zeta \left\vert p^{%
\mathrm{R}}\right\vert \ ^{\zeta }\psi _{n}\left( X\right) \,,\ \ \hat{h}%
_{\perp }^{\mathrm{kin}}\ ^{\zeta }\psi _{n}\left( X\right) =s\pi _{s}\left( 
\mathrm{R}\right) \ ^{\zeta }\psi _{n}\left( X\right) \,,\ \ y\rightarrow
+\infty \,,  \label{2.13}
\end{eqnarray}%
where $\hat{h}_{\perp }^{\mathrm{kin}}=\hat{\Pi}_{z}-\mathbb{I}\left\vert
\mu \right\vert B_{z}\left( y\right) $ is the one-particle transverse
kinetic energy operator. It is important to emphasize both sets of solutions
exist provided the quantum numbers $n$ obey the conditions%
\begin{equation}
\left[ s\pi _{s}\left( \mathrm{L/R}\right) \right] ^{2}>\pi _{x}^{2}\,.
\label{3.5}
\end{equation}%
These inequalities ensure the nontriviality of DP spinors with real
asymptotic momenta $p^{\mathrm{L}}$ and $p^{\mathrm{R}}$ in remote areas  and impart consequences to the quantization of the theory, as shall be discussed below.

At last it should be noted that if the field is homogeneous, then the left\
and right\ asymptotic potentials coincide $U_{\mathrm{R}}=U_{\mathrm{L}%
}\equiv U$ and the step\ is trivial, $\mathbb{U}=0$. As a result, there is
no distinction between the asymptotic momenta $\left\vert p^{\mathrm{L}%
}\right\vert =\left\vert p^{\mathrm{R}}\right\vert =p_{y}=\sqrt{\left(
\omega -sU\right) ^{2}-\pi _{x}^{2}}$ and the transverse kinetic energy
remains the same throughout the space, $\pi _{s}\left( \mathrm{L}\right)
=\pi _{s}\left( \mathrm{R}\right) \equiv \omega _{0}=\omega -sU=\pm \sqrt{%
m^{2}+\mathbf{p}_{\perp }^{2}}$. In a complete absence of external fields, $%
B_{z}\left( y\right) =0\rightarrow U=0$, the transverse kinetic energy $%
\omega _{0}$ fully coincides with the total transverse energy, $\omega
_{0}=\omega =\pm \sqrt{m^{2}+\mathbf{p}_{\perp }^{2}}$.

It should be noted that the time independence of the magnetic field under consideration is an idealization. Physically, it is meaningful to believe that the field inhomogeneity was switched on sufficiently fast before instant $t_{\mathrm{in}}$. By this time, it had time to spread to the whole area under consideration and then acted as a constant field during a large time $T$. It is supposed that one can ignore effects of its switching on and off. This
is a kind of regularization, which could, under certain conditions, be
replaced by periodic boundary conditions in $t$, see Refs. \cite%
{GavGit16,GavGit21}\ for details. Therefore, it is convenient to use the inner
product on the time-like hyperplane $y=\mathrm{const}$, which has the form%
\begin{equation}
\left( \psi ,\psi ^{\prime }\right) _{y}=\int dtdxdz\psi ^{\dagger }\left(
X\right) \gamma ^{0}\gamma ^{2}\psi ^{\prime }\left( X\right) \,,
\label{3.6}
\end{equation}%
after imposing specific normalization conditions\footnote{%
Note that for $\psi ^{\prime }=\psi $, the inner product (\ref{3.6}) divided
by $T$ coincides with the definition of the current density accross the $y$%
-const. hyperplane.}. We assume that all processes take place in a
macroscopically large space-time box, of volume $TV_{y}$, $V_{y}=L_{x}L_{z}$%
, and impose periodic boundary conditions upon DP spinors in the variables $%
t $, $x$, $z$ at the boundaries. Thus, the integrals in (\ref{3.6}) are
calculated from $\left( -T/2,-L_{x}/2,-L_{z}/2\right) $ to $\left(
+T/2,+L_{x}/2,+L_{z}/2\right) $ and the limits $\left( T,L_{x},L_{z}\right)
\rightarrow \infty $ are taken at the end of calculations. The time $T$\ can
be interpreted as a time of observation of the evolution of the system under
consideration. Under these conditions, the inner product is $y$-independent
and we may impose the normalization conditions%
\begin{equation}
\left( \ _{\zeta ^{\prime }}\psi _{n^{\prime }},\ _{\zeta }\psi _{n}\right)
_{y}=\zeta \eta _{\mathrm{L}}\delta _{n^{\prime }n}\delta _{\zeta ^{\prime
}\zeta }\,,\ \ \left( \ ^{\zeta ^{\prime }}\psi _{n^{\prime }},\ ^{\zeta
}\psi _{n}\right) _{y}=\zeta \eta _{\mathrm{R}}\delta _{n^{\prime }n}\delta
_{\zeta ^{\prime }\zeta }\,,\ \ \eta _{\mathrm{L/R}}=\mathrm{sgn}\left[ \pi
_{s}\left( \mathrm{L/R}\right) \right] \,.  \label{3.7}
\end{equation}

Considering that the \textquotedblleft left\textquotedblright\ and
\textquotedblleft right\textquotedblright\ sets of DP spinors are
orthonormal and complete, we may decompose one set into another with the
help of specific coefficients%
\begin{eqnarray}
\eta _{\mathrm{L}}\ ^{\zeta }\psi _{n}\left( X\right) &=&g\left(
_{+}|^{\zeta }\right) \ _{+}\psi _{n}\left( X\right) -g\left( _{-}|^{\zeta
}\right) \ _{-}\psi _{n}\left( X\right) \,,  \notag \\
\eta _{\mathrm{R}}\ _{\zeta }\psi _{n}\left( X\right) &=&g\left(
^{+}|_{\zeta }\right) \ ^{+}\psi _{n}\left( X\right) -g\left( ^{-}|_{\zeta
}\right) \ ^{-}\psi _{n}\left( X\right) \,,  \label{3.9}
\end{eqnarray}%
which, by definition, are inner products between different sets of DP spinors%
\begin{equation}
\left( \ _{\zeta }\psi _{n},\ ^{\zeta ^{\prime }}\psi _{n^{\prime }}\right)
_{y}=\delta _{nn^{\prime }}g\left( _{\zeta }|^{\zeta ^{\prime }}\right)
=\delta _{nn^{\prime }}g\left( ^{\zeta ^{\prime }}|_{\zeta }\right) ^{\ast
}\,.  \label{3.10}
\end{equation}%
Substituting the identities (\ref{3.9}) into the normalization conditions (%
\ref{3.7}) supply us with two important identities%
\begin{equation}
\sum_{\zeta ^{\prime \prime }=\pm }\zeta ^{\prime \prime }g\left( ^{\zeta
^{\prime }}|_{\zeta ^{\prime \prime }}\right) g\left( _{\zeta ^{\prime
\prime }}|^{\zeta }\right) =\zeta \eta _{\mathrm{L}}\eta _{\mathrm{R}}\delta
_{\zeta ^{\prime }\zeta }=\sum_{\zeta ^{\prime \prime }=\pm }\zeta ^{\prime
\prime }g\left( _{\zeta ^{\prime }}|^{\zeta ^{\prime \prime }}\right)
g\left( ^{\zeta ^{\prime \prime }}|_{\zeta }\right) \,,  \label{3.10a}
\end{equation}%
from which we may derive a number of supplementary identities, for example $%
\left\vert g\left( _{+}|^{-}\right) \right\vert ^{2}=\left\vert g\left(
_{-}|^{+}\right) \right\vert ^{2}$, $\left\vert g\left( _{+}|^{+}\right)
\right\vert ^{2}=\left\vert g\left( _{-}|^{-}\right) \right\vert ^{2}$, and $%
\left\vert g\left( _{+}|^{+}\right) \right\vert ^{2}-\left\vert g\left(
_{+}|^{-}\right) \right\vert ^{2}=\eta _{\mathrm{L}}\eta _{\mathrm{R}}$.

The above-presented plane waves can form wave packets in a given asymptotic
region. In this case, the problem can be technically reduced to the problem
of charged-particle creation by an electric step \cite{GavGit16,GavGit21}.
It allows to quantize the DP field operator in the framework of QED with external backgrounds. Based on this quantization, it is possible to calculate all quantities characterizing vacuum instability by steplike magnetic
fields, as discussed in Sec. \ref{Sec3}.

\section{Time-independent Sauter-like magnetic step\label{Sec2b}}

\begin{figure}[th]
\begin{center}
\includegraphics[scale=0.43]{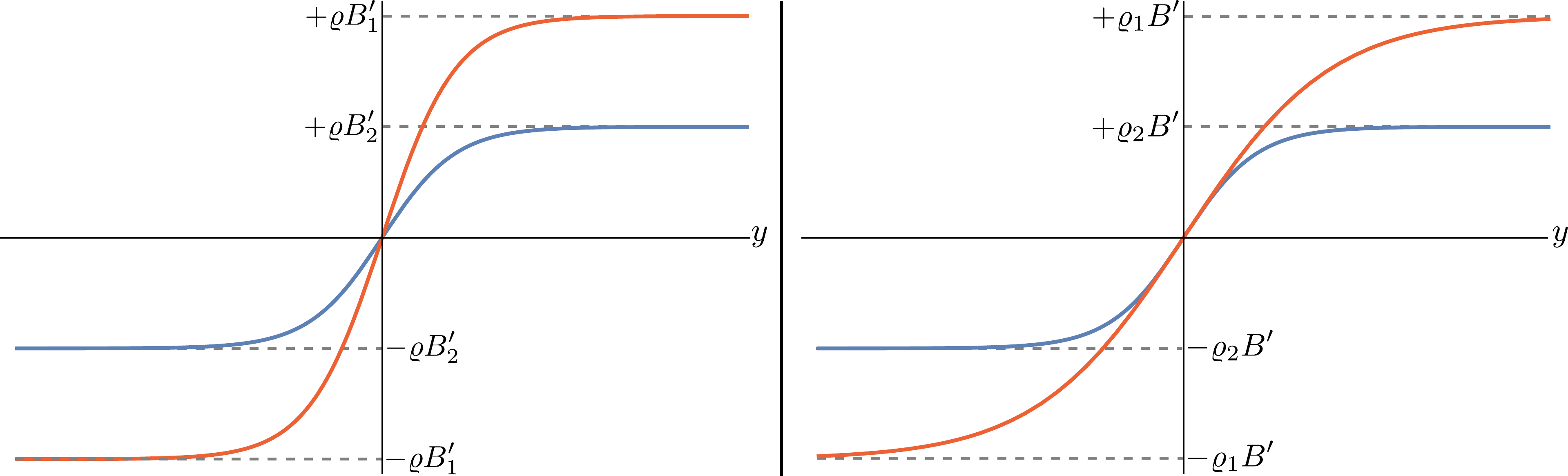}
\end{center}
\caption{Magnetic steps (\protect\ref{2}) as a function of $y$, $\protect%
\varrho $, and $B^{\prime }$. In the left panel, $\protect\varrho $ is the
same for both curves but $B_{1}^{\prime }=2\times B_{2}^{\prime }$. On the
right panel, $B^{\prime }$ is the same for both curves while $\protect%
\varrho _{1}=2\times \protect\varrho _{2}$.}
\label{Fig1}
\end{figure}

To explicitly calculate neutral Fermion pair production, we consider the
following external field%
\begin{equation}
B_{z}\left( y\right) =\varrho B^{\prime }\tanh \left( y/\varrho \right) \,,\
\ B^{\prime }>0\,,\ \ \varrho >0\,,  \label{2}
\end{equation}%
which enjoys the properties discussed before and enables us to solve the DP
equation exactly. More precisely, the field under consideration is
homogeneous at remote distances, $B_{z}\left( \pm \infty \right) =\pm
\varrho B^{\prime }=\mathrm{const.}$, and its gradient is always positive $%
\partial _{y}B_{z}\left( y\right) =B^{\prime }\cosh ^{-2}\left( y/\varrho
\right) \geq 0$; in particular, $\mathbb{U}=2\varrho \left\vert \mu
\right\vert B^{\prime }$ for this field. The amplitude\footnote{%
We represent $B^{\prime }$ with a \textquotedblleft prime\textquotedblright\
to emphasize that it corresponds to the amplitude of the gradient, $\left.
\partial _{y}B_{z}\left( y\right) \right\vert _{y=0}=B^{\prime }$, rather
than the field (\ref{2}), which is intrinsically a \textquotedblleft
step\textquotedblright\ and not a \textquotedblleft
barrier\textquotedblright .} $B^{\prime }$ and the \textquotedblleft
inhomogeneity length\textquotedblright\ $\varrho $ describe, respectively,
the \textquotedblleft slope\textquotedblright\ of the field with respect to
the $y$-axis and how \textquotedblleft rectilinear\textquotedblright\ it is
in the neighborhood of the origin. Thus, the larger $B^{\prime }$ and $%
\varrho $, the more \textquotedblleft steep\textquotedblright\ and the more
\textquotedblleft rectilinear\textquotedblright\ the pattern of (\ref{2})
near the origin. For illustrative purposes, we present in Fig. \ref{Fig1}
the magnetic step (\ref{2}) as a function of $y$, $\varrho $, and $B^{\prime
}$. It should be noted that the field (\ref{2}) shares the same functional
form as the time-independent scalar electric potential originally considered
by F. Sauter in \cite{Sauter31}, who first solved the Dirac equation with
this field. 

Inserting the external field (\ref{2}) into Eq. (\ref{2.11}) and performing
a simultaneous change of variables%
\begin{equation}
\varphi _{n,\chi }\left( y\right) =\xi ^{\rho }\left( 1-\xi \right) ^{\sigma
}f\left( \xi \right) \,,\ \ \xi \left( y\right) =\frac{1}{2}\left[ 1+\tanh
\left( y/\varrho \right) \right] \,,  \label{3.1}
\end{equation}%
we may convert Eq. (\ref{2.11}) to the same\ form as the differential
equation for the Gauss Hypergeometric Function \cite{Erdelyivol1}%
\begin{equation}
\xi \left( 1-\xi \right) f^{\prime \prime }+\left[ c-\left( a+b+1\right) \xi %
\right] f^{\prime }-abf=0\,,  \label{3.1b}
\end{equation}%
provided the parameters $\rho $, $\sigma $, $a$, $b$, and $c$ are:%
\begin{eqnarray}
a &=&\frac{1}{2}\left( 1-\chi \right) -\frac{i\varrho }{2}\left( \mathbb{U}%
+\left\vert p^{\mathrm{L}}\right\vert -\left\vert p^{\mathrm{R}}\right\vert
\right) \,,  \notag \\
b &=&\frac{1}{2}\left( 1+\chi \right) +\frac{i\varrho }{2}\left( \mathbb{U}%
+\left\vert p^{\mathrm{R}}\right\vert -\left\vert p^{\mathrm{L}}\right\vert
\right) \,,  \notag \\
c &=&1-i\varrho \left\vert p^{\mathrm{L}}\right\vert \,,\ \ \rho =-\frac{%
i\varrho }{2}\left\vert p^{\mathrm{L}}\right\vert \,,\ \ \sigma =\frac{%
i\varrho }{2}\left\vert p^{\mathrm{R}}\right\vert \,.  \label{3.2}
\end{eqnarray}

Among the 24 Hypergeometric functions satisfying Eq. (\ref{3.1b}) \cite%
{Erdelyivol1}, we select those that tend to unity as $y\rightarrow \mp
\infty $. Solutions meeting this property are proportional to Hypergeometric
functions of type $F\left( a^{\prime },b^{\prime };c^{\prime };\xi \right) $
and $F\left( a^{\prime \prime },b^{\prime \prime };c^{\prime \prime };1-\xi
\right) $. For example, a possible set of solutions to Eq. (\ref{2.11})
behaving asymptotically like Eqs. (\ref{2.14}) is%
\begin{eqnarray}
\ _{\zeta }\varphi _{n,\chi }\left( y\right) &=&\ _{\zeta }\mathcal{N}\exp
\left( i\zeta \left\vert p^{\mathrm{L}}\right\vert y\right) \left[ 1+\exp
\left( 2y/\varrho \right) \right] ^{-i\varrho \left( \zeta \left\vert p^{%
\mathrm{L}}\right\vert +\left\vert p^{\mathrm{R}}\right\vert \right) /2}\
_{\zeta }u\left( \xi \right) \,,  \notag \\
\ ^{\zeta }\varphi _{n,\chi }\left( y\right) &=&\ ^{\zeta }\mathcal{N}\exp
\left( i\zeta \left\vert p^{\mathrm{R}}\right\vert y\right) \left[ 1+\exp
\left( -2y/\varrho \right) \right] ^{i\varrho \left( \left\vert p^{\mathrm{L}%
}\right\vert +\zeta \left\vert p^{\mathrm{R}}\right\vert \right) /2}\
^{\zeta }u\left( \xi \right) \,,  \label{3.3}
\end{eqnarray}%
where%
\begin{eqnarray}
\ _{-}u\left( \xi \right) &=&F\left( a,b;c;\xi \right) \,,\ \ _{+}u\left(
\xi \right) =F\left( a+1-c,b+1-c;2-c;\xi \right) \,,  \notag \\
\ ^{-}u\left( \xi \right) &=&F\left( a,b;a+b+1-c;1-\xi \right) \,,\
^{+}u\left( \xi \right) =F\left( c-a,c-b;c+1-a-b;1-\xi \right) \,.
\label{3.4}
\end{eqnarray}%
With the aid of these solutions, we may finally introduce the sets of DP
spinors%
\begin{eqnarray}
\ _{\zeta }\psi _{n}\left( X\right) &=&e^{-i\left(
p_{0}t-p_{x}x-p_{z}z\right) }\left( \mathbb{I}+sR\right) \left[ \hat{\pi}%
_{z}+\mathbb{I}\left( -\left\vert \mu \right\vert B_{z}\left( y\right)
+s\omega \right) \right] \ _{\zeta }\varphi _{n,\chi }\left( y\right)
\upsilon _{\kappa }^{\left( \chi \right) }\,,  \notag \\
\ ^{\zeta }\psi _{n}\left( X\right) &=&e^{-i\left(
p_{0}t-p_{x}x-p_{z}z\right) }\left( \mathbb{I}+sR\right) \left[ \hat{\pi}%
_{z}+\mathbb{I}\left( -\left\vert \mu \right\vert B_{z}\left( y\right)
+s\omega \right) \right] \ ^{\zeta }\varphi _{n,\chi }\left( y\right)
\upsilon _{\kappa }^{\left( \chi \right) }\,,  \label{3.4b}
\end{eqnarray}%
provided the quantum numbers $n$ obey the restrictions given by Eq. (\ref%
{3.5}). Under these conditions, we may calculate the normalization constants$%
\ _{\zeta }\mathcal{N}$,$\ ^{\zeta }\mathcal{N}$ and present the final form
of DP spinors. Since the inner product (\ref{3.6}) is $y$-independent, we
may calculate the normalization constants at remote regions $y\rightarrow
\mp \infty $, where the DP spinors admit the asymptotic forms (\ref{2.16}).
Evaluating the inner product (\ref{3.6}) and imposing the normalization
conditions (\ref{3.7}), we finally obtain%
\begin{equation}
\left\vert \ _{\zeta }\mathcal{N}\right\vert =\frac{\left[ TV_{y}\Upsilon
\left( 1-s\kappa \chi \Upsilon \right) \right] ^{-1/2}}{2\sqrt{\left\vert p^{%
\mathrm{L}}\right\vert \left\vert \pi _{s}\left( \mathrm{L}\right) -s\chi
\zeta \left\vert p^{\mathrm{L}}\right\vert \right\vert }}\,,\ \ \left\vert \
^{\zeta }\mathcal{N}\right\vert =\frac{\left[ TV_{y}\Upsilon \left(
1-s\kappa \chi \Upsilon \right) \right] ^{-1/2}}{2\sqrt{\left\vert p^{%
\mathrm{R}}\right\vert \left\vert \pi _{s}\left( \mathrm{R}\right) -s\chi
\zeta \left\vert p^{\mathrm{R}}\right\vert \right\vert }}\,.  \label{3.8}
\end{equation}

\section{Vacuum instability processes\label{Sec3}}

As discussed in the preceding section, both the nontriviality of DP-spinors
as the assumption on their completeness (\ref{3.9}) crucially depends on the
restrictions imposed by Eq. (\ref{3.5}). This inequality imposes certain
limitations upon the quantum numbers. For example, for critical external
fields, whose step magnitude obeys the inequality%
\begin{equation}
\mathbb{U}>\mathbb{U}_{\mathrm{c}}=2m\,,  \label{3.10b}
\end{equation}%
the whole manifold of quantum numbers can be divided into five sub-ranges, $%
\Omega _{k}\,,\ k=1,...,5$. While the division of each sub-range $\Omega
_{k} $ can be realized by following the same considerations developed for
charged particles in\ an electric step \cite{GavGit16}, here we stick to the
range where particle creation is possible, the so-called Klein zone $\Omega
_{3}$. This sub-range, which exists only for critical fields (\ref{3.10b}),
is defined by a bounded set of quantum numbers%
\begin{equation}
\Omega _{3}=\left\{ n:U_{\mathrm{L}}+\pi _{x}\leq s\omega \leq U_{\mathrm{R}%
}-\pi _{x}\,,\ \ \pi _{xz}\leq \mathbb{U}/2\right\} \,,\ \ \pi _{xz}=\sqrt{%
\pi _{x}^{2}+p_{z}^{2}}\,,  \label{3.11}
\end{equation}%
for which the restrictions $s\pi _{s}\left( \mathrm{L}\right) \geq \pi _{x}$
and $s\pi _{s}\left( \mathrm{R}\right) \leq -\pi _{x}$ are satisfied. In
particular, $s\eta _{\mathrm{L}}=+1$ while $s\eta _{\mathrm{R}}=-1$. As a
result, two linearly-independent \textquotedblleft left\textquotedblright\ $%
\ _{\zeta }\psi _{n_{3}}\left( X\right) $ and \textquotedblleft
right\textquotedblright\ $\ ^{\zeta }\psi _{n_{3}}\left( X\right) $ sets of
DP spinors do exist for quantum numbers within the Klein zone, $n_{3}=n\in
\Omega _{3}$.

The quantization of DP fields is realized using sets of exact solutions with
special properties in remote areas. More specifically, one needs to classify
stationary solutions as particle or antiparticle states and as incoming
waves (waves traveling toward the \textquotedblleft step\textquotedblright )
or outgoing waves (waves traveling outward the \textquotedblleft
step\textquotedblright ) in remote areas. Selecting such solutions demands
careful consideration of the inner product between DP spinors on $y$- and $t$%
-constant hyperplanes because important quantities to the scattering problem
are expressed as surface integrals on such hyperplanes. Examples include
classical/quantum kinetic energies, current field operators, magnetic moment
field operators, and fluxes of particle/antiparticle energies across $y$%
-constant hyperplanes in remote areas. A detailed study of these quantities
was presented in Refs. \cite{GavGit16,GavGit21} for charged particles and in 
\cite{GavGit13} for neutral Fermions. For Fermions with $s=+1$, it was
demonstrated that the set $\ _{\zeta }\psi _{n_{3}}\left( X\right) $
corresponds to antiparticle states while the set $\ ^{\zeta }\psi
_{n_{3}}\left( X\right) $ corresponds to particle states in specific remote
areas. Moreover, \textquotedblleft in\textquotedblright -solutions (incoming
waves) and \textquotedblleft out\textquotedblright -solutions (outgoing
waves) are:%
\begin{equation}
\text{\textrm{in-solutions:}\ }\ _{-}\psi _{n}\left( X\right) \,,\ ^{-}\psi
_{n}\left( X\right) \,,\ \ \text{\textrm{out-solutions:}\ }\ _{+}\psi
_{n}\left( X\right) \,,\ ^{+}\psi _{n}\left( X\right) \,,\ \ n\in \Omega
_{3}\,.  \label{3.12}
\end{equation}%
The above sets of solutions are complete and orthogonal with respect to the
inner product on $t$-constant hyperplane%
\begin{equation}
\left( \psi _{n},\psi _{n^{\prime }}^{\prime }\right)
=\int_{V_{y}}dxdz\int_{-K^{\left( \mathrm{L}\right) }}^{K^{\left( \mathrm{R}%
\right) }}dy\psi _{n}^{\dagger }\left( X\right) \psi _{n^{\prime }}^{\prime
}\left( X\right) \,,  \label{3.13}
\end{equation}%
where the lower/upper cutoffs $K^{\left( \mathrm{L/R}\right) }$ are
macroscopic but finite parameters of the volume regularization that are
situated far beyond the region of a large gradient $\partial _{y}B_{z}\left(
y\right) $. The cutoffs are chosen so that the principal value of integral (%
\ref{3.13}) is determined by integrals over areas where the gradient $%
\partial _{y}B_{z}\left( y\right) $ is small, and its influence on pairs
creation can be neglected. Macroscopic times of motion of particles and
antiparticles, $t^{\left( \mathrm{L}\right) }=K^{\left( \mathrm{L}\right)
}\left\vert \pi _{s}\left( \mathrm{L}\right) /p^{\mathrm{L}}\right\vert $\
and $t^{\left( \mathrm{R}\right) }=K^{\left( \mathrm{R}\right) }\left\vert
\pi _{s}\left( \mathrm{R}\right) /p^{\mathrm{R}}\right\vert $, in the \
regions of a small gradient are assumed to obey the condition%
\begin{equation}
t^{\left( \mathrm{L}\right) }-t^{\left( \mathrm{R}\right) }=O\left( 1\right)
,  \label{i8}
\end{equation}%
where $O\left( 1\right) $ denotes terms that are negligibly small compared
to $t^{\left(\mathrm{L/R}\right)}$. The inner product (\ref{3.13}) between
both sets of DP spinors reads: 
\begin{equation}
\left( \ _{\zeta }\psi _{n},\ _{\zeta }\psi _{n^{\prime }}\right) =\left( \
^{\zeta }\psi _{n},\ ^{\zeta }\psi _{n^{\prime }}\right) =\mathcal{M}%
_{n}\delta _{nn^{\prime }}\,,\ \ \left( \ _{\zeta }\psi _{n},\ ^{\zeta }\psi
_{n^{\prime }}\right) =0\,,\ \ n,n^{\prime }\in \Omega _{3}\,,  \label{3.14}
\end{equation}%
where $\mathcal{M}_{n}=2\left\vert g\left( _{+}|^{-}\right) \right\vert
^{2}t^{\left( \mathrm{L/R}\right) }/T$; see Ref. \cite{GavGit13,GavGit16}.
Due to classification (\ref{3.12}) and the above properties, we may quantize
the DP field operator $\hat{\Psi}\left( X\right) $ by decomposing it in ($y$%
-independent) sets of creation and annihilation operators of particles and
antiparticles. Because there are two linearly independent sets of spinors (%
\ref{3.12}), the quantization is performed using two distinct
\textquotedblleft in\textquotedblright\ and \textquotedblleft
out\textquotedblright\ sets of annihilation \& creation operators%
\begin{eqnarray}
&&\text{\textrm{in-set:}}\ _{-}b_{n_{3}}\left( \mathrm{in}\right) \,,\
_{-}b_{n_{3}}^{\dagger }\left( \mathrm{in}\right) \,,\ ^{-}a_{n_{3}}\left( 
\mathrm{in}\right) \,,\ ^{-}a_{n_{3}}^{\dagger }\left( \mathrm{in}\right) \,,
\notag \\
&&\text{\textrm{out-set:}}\ _{+}b_{n_{3}}\left( \mathrm{out}\right) \,,\
_{+}b_{n_{3}}^{\dagger }\left( \mathrm{out}\right) \,,\ ^{+}a_{n_{3}}\left( 
\mathrm{out}\right) \,,\ ^{+}a_{n_{3}}^{\dagger }\left( \mathrm{out}\right)
\,,  \label{3.15}
\end{eqnarray}%
which, in turn, obey the following anticommutation relations%
\begin{eqnarray}
&&\left[ \ ^{-}a_{n_{3}^{\prime }}\left( \mathrm{in}\right) ,\
^{-}a_{n_{3}}^{\dagger }\left( \mathrm{in}\right) \right] _{+}=\left[ \
_{-}b_{n_{3}^{\prime }}\left( \mathrm{in}\right) ,\ _{-}b_{n_{3}}^{\dagger
}\left( \mathrm{in}\right) \right] _{+}=\delta _{n_{3}^{\prime }n_{3}}\,, 
\notag \\
&&\left[ \ ^{+}a_{n_{3}^{\prime }}\left( \mathrm{out}\right) ,\
^{+}a_{n_{3}}^{\dagger }\left( \mathrm{out}\right) \right] _{+}=\left[ \
_{+}b_{n_{3}^{\prime }}\left( \mathrm{out}\right) ,\ _{+}b_{n_{3}}^{\dagger
}\left( \mathrm{out}\right) \right] _{+}=\delta _{n_{3}^{\prime }n_{3}}\,,
\label{3.16}
\end{eqnarray}%
and whose annihilation operators (\ref{3.15}) annihilate the corresponding
vacuum states\footnote{%
Rigorously, there are five \textquotedblleft in\textquotedblright\ vacuum
states $\left\vert 0,\mathrm{in}\right\rangle ^{\left( i\right) }$ and five
\textquotedblleft out\textquotedblright\ vacuum states $\left\vert 0,\mathrm{%
out}\right\rangle ^{\left( i\right) }$, each corresponding to vacuum states
for quantum numbers defined within the five existing subranges $n_{i}\in
\Omega _{i}$, $i=1,...,5$ \cite{GavGit16}. Because we are restricting to
processes within the Klein zone, we omit the superscript $\left( 3\right) $
on the partial vacua (\ref{3.17}) for simplicity.}%
\begin{equation}
\ _{-}b_{n_{3}}\left( \mathrm{in}\right) \left\vert 0,\mathrm{in}%
\right\rangle =\ ^{-}a_{n_{3}}\left( \mathrm{in}\right) \left\vert 0,\mathrm{%
in}\right\rangle =0\,,\ _{+}b_{n_{3}}\left( \mathrm{out}\right) \left\vert 0,%
\mathrm{out}\right\rangle =\ ^{+}a_{n_{3}}\left( \mathrm{out}\right)
\left\vert 0,\mathrm{out}\right\rangle =0\,.  \label{3.17}
\end{equation}%
The algebra (\ref{3.16}) realizes the equal-time anticommutation relations
for DP Fermion fields in the Heisenberg representation \cite{GavGit13}.
Finally, the quantized DP field operator in the Klein zone reads%
\begin{eqnarray}
\hat{\Psi}\left( X\right) &=&\sum_{n\in \Omega _{3}}\mathcal{M}_{n}^{-1/2}%
\left[ \ ^{-}a_{n}\left( \mathrm{in}\right) \ ^{-}\psi _{n}\left( X\right)
+\ _{-}b_{n}^{\dagger }\left( \mathrm{in}\right) \ _{-}\psi _{n}\left(
X\right) \right] \,,  \notag \\
&=&\sum_{n\in \Omega _{3}}\mathcal{M}_{n}^{-1/2}\left[ \ ^{+}a_{n}\left( 
\mathrm{out}\right) \ ^{+}\psi _{n}\left( X\right) +\ _{+}b_{n}^{\dagger
}\left( \mathrm{out}\right) \ _{+}\psi _{n}\left( X\right) \right] \,.
\label{3.18}
\end{eqnarray}

Using orthogonality relations between DP spinors (\ref{3.7}), (\ref{3.14})
and the relations given by Eqs. (\ref{3.9}), we may find a linear relation
between the \textquotedblleft in\textquotedblright -set of
creation/annihilation operators in terms of the \textquotedblleft
out\textquotedblright -set and \textit{vice-versa}. For example, two (out of
four) canonical transformations have the following form%
\begin{eqnarray}
&&\ _{-}b_{n}^{\dagger }\left( \mathrm{in}\right) =-g\left( ^{+}|_{-}\right)
^{-1}\ ^{+}a_{n}\left( \mathrm{out}\right) +g\left( _{+}|^{-}\right)
^{-1}g\left( _{-}|^{-}\right) \ _{+}b_{n}^{\dagger }\left( \mathrm{out}%
\right) \,,  \notag \\
&&\ ^{+}a_{n}\left( \mathrm{out}\right) =-g\left( _{-}|^{+}\right) ^{-1}\
_{-}b_{n}^{\dagger }\left( \mathrm{in}\right) +g\left( ^{-}|_{+}\right)
^{-1}g\left( ^{+}|_{+}\right) \ ^{-}a_{n}\left( \mathrm{in}\right) \,.
\label{3.19}
\end{eqnarray}%
With the aid of the canonical transformations (\ref{3.19}), we may finally
compute important quantities to the study of pair creation, such as the
differential mean numbers of \textquotedblleft out\textquotedblright\
particles created from the \textquotedblleft in\textquotedblright\ vacuum,%
\begin{equation}
N_{n}^{\mathrm{cr}}=\left\langle 0,\mathrm{in}\left\vert \
^{+}a_{n}^{\dagger }\left( \mathrm{out}\right) \ ^{+}a_{n}\left( \mathrm{out}%
\right) \right\vert \mathrm{in},0\right\rangle =\left\vert g\left(
_{-}|^{+}\right) \right\vert ^{-2}\,,\ \ n\in \Omega _{3}\,,  \label{3.20}
\end{equation}%
and the flux density of particles created with a given\emph{\ }$s$,%
\begin{equation}
n_{s}^{\mathrm{cr}}=\frac{1}{V_{y}T}\sum_{n\in \Omega _{3}}N_{n}^{\mathrm{cr}%
}=\frac{1}{\left( 2\pi \right) ^{3}}\int dp_{z}\int dp_{x}\int dp_{0}N_{n}^{%
\mathrm{cr}}\,.  \label{3.21}
\end{equation}%
It should be noted that $n_{+1}^{\mathrm{cr}}=n_{-1}^{\mathrm{cr}}$. The total flux
density of particles created with both\emph{\ }$s=\pm 1$\emph{\ }is $n^{%
\mathrm{cr}}\mathcal{=}n_{+1}^{\mathrm{cr}}+n_{-1}^{\mathrm{cr}}$ and the
vacuum-vacuum transition probability reads:%
\begin{equation}
P_{v}=\left\vert \left\langle 0,\mathrm{out}|0,\mathrm{in}\right\rangle
\right\vert ^{2}=\exp \left[ \sum_{s=\pm 1}\sum_{n\in \Omega _{3}}\ln \left(
1-N_{n}^{\mathrm{cr}}\right) \right] \,.  \label{3.22}
\end{equation}

As mentioned in the preceding section, the mean numbers of
\textquotedblleft out\textquotedblright\ antiparticles created from the
\textquotedblleft in\textquotedblright\ vacuum $N_{n}^{\mathrm{cr}%
}=\left\langle 0,\mathrm{in}\left\vert \ _{+}b_{n}^{\dagger }\left( \mathrm{%
out}\right) \ _{+}b_{n}\left( \mathrm{out}\right) \right\vert \mathrm{in}%
,0\right\rangle =\left\vert g\left( _{+}|^{-}\right) \right\vert ^{-2}$ are
given by Eq. (\ref{3.20}) due to the identity $\left\vert g\left(
_{+}|^{-}\right) \right\vert ^{-2}=\left\vert g\left( _{-}|^{+}\right)
\right\vert ^{-2}$. To obtain the rightmost expression in Eq. (\ref{3.22}),
one needs to find an unitary operator $V_{\Omega _{3}}$ that connects the
\textquotedblleft in\textquotedblright\ and \textquotedblleft
out\textquotedblright\ vacua, $\left\vert 0,\mathrm{out}\right\rangle
=V_{\Omega _{3}}^{\dagger }\left\vert 0,\mathrm{in}\right\rangle $; see e.g.
Refs. \cite{GavGit16,AdoGavGit20} for its explicit form.

If the total number of created particles\emph{\ }$N^{\mathrm{cr}}=V_{y}Tn^{%
\mathrm{cr}}$ is small, one may neglect higher-order terms in Eq. (\ref{3.22}%
) to conclude that the vacuum-vacuum transition probability slightly
deviates from the unity, $P_{v}\approx 1-N^{\mathrm{cr}}$. This indicates
that the external field weakly violates the vacuum. Assuming that
the effective action $S_{\mathrm{eff}}$\ for this problem satisfies
the Schwinger relation $P_{v}=\exp \left( -2\mathrm{Im}S_{\mathrm{eff%
}}\right) $, we may straightforwardly establish a connection between the
effective action with the flux density (\ref{3.21}) by taking into
account that its imaginary part is also small in this regime, $%
P_{v}\approx 1-2\mathrm{Im}S_{\mathrm{eff}}$. Therefore,%
\begin{equation}
\mathrm{Im}S_{\mathrm{eff}}\approx V_{y}Tn^{\mathrm{cr}}/2\,.  \label{3.39}
\end{equation}

According to Eqs. (\ref{3.20}) - (\ref{3.22}), all the information about
pair creation by the external field is enclosed in $g\left( _{-}|^{+}\right) 
$. To obtain this coefficient, we may use an appropriate Kummer \cite%
{Erdelyivol1} relation that connects three Gauss Hypergeometric functions
appearing in one of the relations given by Eq. (\ref{3.9}). After
straightfoward calculations, we discover that%
\begin{equation}
g\left( _{-}|^{+}\right) =-\eta _{\mathrm{L}}\frac{\ ^{+}\mathcal{N}\Gamma
\left( c-a-b+1\right) \Gamma \left( 1-c\right) }{\ _{-}\mathcal{N}\Gamma
\left( 1-a\right) \Gamma \left( 1-b\right) }\,.  \label{3.23}
\end{equation}%
Calculating the absolute square $\left\vert g\left( _{-}|^{+}\right)
\right\vert ^{-2}$, we finally obtain the differential mean number of pairs
created from the vacuum:%
\begin{equation}
N_{n}^{\mathrm{cr}}=\frac{\sinh \left( \pi \varrho \left\vert p^{\mathrm{R}%
}\right\vert \right) \sinh \left( \pi \varrho \left\vert p^{\mathrm{L}%
}\right\vert \right) }{\sinh \left[ \pi \varrho \left( \mathbb{U}+\left\vert
p^{\mathrm{L}}\right\vert -\left\vert p^{\mathrm{R}}\right\vert \right) /2%
\right] \sinh \left[ \pi \varrho \left( \mathbb{U}+\left\vert p^{\mathrm{R}%
}\right\vert -\left\vert p^{\mathrm{L}}\right\vert \right) /2\right] }\,.
\label{3.24}
\end{equation}%
Note that $N_{n}^{\mathrm{cr}}$ is positive-definite because the difference $%
\left\vert \left\vert p^{\mathrm{L}}\right\vert -\left\vert p^{\mathrm{R}%
}\right\vert \right\vert $ bounded in this subrange; $0\leq \left\vert
\left\vert p^{\mathrm{L}}\right\vert -\left\vert p^{\mathrm{R}}\right\vert
\right\vert \leq \sqrt{\mathbb{U}\left( \mathbb{U}-2\pi _{x}\right) }$. The
above expression gives the exact distribution of neutral Fermions created
from the vacuum by the field (\ref{2}).\ When summed over the quantum
numbers, it provides exact expressions for the flux density of the created
particles (\ref{3.21}) and the vacuum-vacuum transition probability (\ref%
{3.22}). Lastly, it is noteworthy to discuss some peculiarities associated
with the choice of the quantum number $s$ and its impact on the quantization
(\ref{3.18}). As pointed out in Sec. \ref{Sec2}, there are two species of
neutral Fermions, one with $s=+1$ and another with $s=-1$. In the latter
case, the classification differs from the one given by Eq (\ref{3.12}),
namely $\ _{+}\psi _{n_{3}}\left( X\right) \,,\ ^{+}\psi _{n_{3}}\left(
X\right) $ are \textquotedblleft in\textquotedblright -solutions while $\
_{-}\psi _{n_{3}}\left( X\right) \,,\ ^{-}\psi _{n_{3}}\left( X\right) $ are
\textquotedblleft out\textquotedblright -solutions. Although this
classification changes the quantization (\ref{3.18}), it does not change the
mean number (\ref{3.20}). This means that the summations over $s$ in Eqs. (%
\ref{3.21}), (\ref{3.22}) just produce an extra factor of $2$ in final
expressions. That is why it is enough choosing $s$ fixed to perform specific
calculations; hereafter, we select $s=+1$ for convenience.

\section{Pair creation in special configurations\label{Sec3.1}}

To unveil important features about pair creation, here we study the
differential and total quantities introduced before in situations where the
external field lies in two special configurations, namely when the field
\textquotedblleft gradually\textquotedblright\ varies along the $y$-axis and
\textquotedblleft sharply\textquotedblright\ varies near the origin $y=0$.
For convenience, we separately discuss each configuration below.

\subsection{\textquotedblleft \textit{Gradually}\textquotedblright -varying
field configuration\label{Sec3.1.1}}

This field configuration corresponds to the case where the amplitude $%
B^{\prime }$ is sufficiently large and the field inhomogeneity stretches
over a relatively wide region of the space, such that the condition%
\begin{equation}
\sqrt{\varrho \mathbb{U}/2}\gg \max \left( 1,\frac{m}{\sqrt{\left\vert \mu
\right\vert B^{\prime }}}\right) \,,  \label{3.25}
\end{equation}%
is satisfied. Accordingly, the arguments of the hyperbolic functions in (\ref%
{3.24}) are large, meaning that the mean number of pairs created acquires
the following approximate form,%
\begin{equation}
N_{n}^{\mathrm{cr}}\approx e^{-\pi \tau }\,,\ \ \tau =\varrho \left( \mathbb{%
U}-\left\vert p^{\mathrm{R}}\right\vert -\left\vert p^{\mathrm{L}%
}\right\vert \right) \,.  \label{3.26}
\end{equation}%
Let us study the behavior of the approximation (\ref{3.26}) with respect to
the quantum numbers. According to Eq. (\ref{2.15}), $\tau $ grows
monotonically with $\omega $ and $p_{x}$, which means that it has a minimum
at $\omega =p_{x}=0$. At this point, $\tau =\tau _{0}=\left. \tau
\right\vert _{\omega =p_{x}=0}\approx m^{2}/\left\vert \mu \right\vert
B^{\prime }$, and the distribution (\ref{3.26}) reaches its maximum, $N_{n}^{%
\mathrm{cr}}\approx N_{n}^{\max }=\exp \left( -\pi m^{2}/\left\vert \mu
\right\vert B^{\prime }\right) $. If $\omega =0$ but $\left\vert
p_{x}\right\vert $ deviates from the origin such that it remains
sufficiently away from the borders of the Klein zone (\ref{3.11}), $\tau $
is approximately given by%
\begin{equation}
\tau =\lambda +O\left( \varrho \pi _{x}^{4}/\mathbb{U}^{3}\right) \,,\ \
\lambda =\frac{\pi _{x}^{2}}{\left\vert \mu \right\vert B^{\prime }}\,,
\label{3.27}
\end{equation}%
and the mean number (\ref{3.26}) approaches to the uniform distribution, $%
N_{n}^{\mathrm{uni}}=\exp \left( -\pi \lambda \right) $, found earlier in
Ref. \cite{GavGit13} for the case of a linearly growing magnetic field. This
similarity is not unexpected because the field profile approaches a linearly
growing magnetic field at regions sufficiently near the origin as soon as $%
\varrho $ increases. In other words, for sufficiently large $\varrho $, the
gradient of the magnetic field (\ref{2}) becomes almost constant and that is
why the differential mean number of pairs created by this field tends to the
uniform distribution in the regime (\ref{3.25}). However, this similarity is
just local as the distribution $N_{n}^{\mathrm{uni}}$ cannot be uniformly
extended to the whole Klein zone. For example, let us analyze cases where
either $p_{x}$ or $\omega $ are sufficiently large. According to the
conditions (\ref{3.11}), for values of $p_{x}$ close enough to the borders
of the Klein zone, $p_{x}^{2}\lesssim \left( \mathbb{U}/2\right) ^{2}-m^{2}$%
, we observe that $\omega \approx 0$ and both momenta $\left\vert p^{\mathrm{%
L}}\right\vert $, $\left\vert p^{\mathrm{R}}\right\vert $ are significantly
small. As a result, the mean number (\ref{3.26}) is exponentially small in
this case $N_{n}^{\mathrm{cr}}\approx \exp \left( -\pi \varrho \mathbb{U}%
\right) $ (thus, quite distinct of $N_{n}^{\mathrm{uni}}$). In the oposite
situation, that is if $\left\vert \omega \right\vert \lesssim \mathbb{U}%
/2-\pi _{x}$, we see that either $\left\vert p^{\mathrm{L}}\right\vert $ or $%
\left\vert p^{\mathrm{R}}\right\vert $ approaches its maximum value $\sqrt{%
\mathbb{U}\left( \mathbb{U}-2\pi _{x}\right) }$ while the remaining one
tends to zero. For example, if $\omega $ is large and positive, say $\omega =%
\mathbb{U}/2-\pi _{x}-0^{+}$, then $\left\vert p^{\mathrm{L}}\right\vert =%
\mathbb{U}\left[ 1+O\left( \pi _{x}/\mathbb{U}\right) \right] $ while $%
\left\vert p^{\mathrm{R}}\right\vert \approx 0$. In this case, the mean
number (\ref{3.26}) is also small due to the condition (\ref{3.25}) and,
again, quite different from the uniform distribution $N_{n}^{\mathrm{uni}}$.
Hence, we observe that the most significant contribution to (\ref{3.26})
comes from finite values of $\left\vert p_{x}\right\vert $ and from a
relatively wide range of $\omega $ but, still, sufficiently away from the
borders of the Klein zone (\ref{3.11}) such that the conditions $\min \left(
\pi _{+1}^{2}\left( \mathrm{L}\right) ,\pi _{+1}^{2}\left( \mathrm{R}\right)
\right) \gg \pi _{x}^{2}$ remains valid. In this case, $\tau $ admits the
following approximation%
\begin{equation}
\tau =\frac{\left( \mathbb{U}/2\right) ^{2}}{\left( \mathbb{U}/2\right)
^{2}-\omega ^{2}}\lambda +O\left( \pi _{x}^{4}/\left\vert \pi _{+1}\left( 
\mathrm{R}\right) \right\vert ^{3}\right) +O\left( \pi _{x}^{4}/\left\vert
\pi _{+1}\left( \mathrm{L}\right) \right\vert ^{3}\right) \,.  \label{3.28}
\end{equation}

Now, we can estimate the flux density of pairs created $n^{\mathrm{cr}}$\
for a magnetic step evolving gradually along the $y$-direction according to (%
\ref{3.25}). To this end, it is convenient to transform the original
integral over $p_{0}$ into an integral over $\omega $ through the relation
between $p_{0}$, $\omega $, and $p_{z}$ discussed before, $p_{0}^{2}=\omega
^{2}+p_{z}^{2}$. Performing such a change of variables, the flux density of
the particles created by the external field in the configuration (\ref{3.25}%
) has the form%
\begin{eqnarray}
&&n^{\mathrm{cr}}\approx \frac{4}{\left( 2\pi \right) ^{3}}%
\int_{0}^{p_{z}^{\max }}dp_{z}\int_{-p_{x}^{\max }}^{p_{x}^{\max
}}dp_{x}\int_{0}^{\omega _{\max }^{2}}d\omega ^{2}\frac{e^{-\pi \tau }}{%
\sqrt{\omega ^{2}+p_{z}^{2}}}\,,  \notag \\
&&p_{z}^{\max }=\sqrt{\left( \mathbb{U}/2\right) ^{2}-m^{2}}\,,\ \
p_{x}^{\max }=\sqrt{\left( \mathbb{U}/2\right) ^{2}-m^{2}-p_{z}^{2}}\,,\ \
\omega _{\max }=\mathbb{U}/2-\pi _{x}\,.  \label{3.29}
\end{eqnarray}%
The multiplicative factor $4$ comes from the summation over $s$ and from the
fact that the integrand is symmetric in $p_{z}$. To obtain an analytical
expression to $N^{\mathrm{cr}}$, we formally extend the integration limits
of the last two integrals to infinity. This procedure amounts to
incorporating exponentially small contributions to $n^{\mathrm{cr}}$ since
the differential mean number is exponentially small at large $p_{x}$ and $%
\omega $. In this case, we may technically interchange the order of the last
two integrals in (\ref{3.29}) and use the approximation given by Eq. (\ref%
{3.28}) to discover that the flux density of the created particles is
approximately given by%
\begin{equation}
n^{\mathrm{cr}}\approx \frac{1}{2\pi ^{3}}\varrho ^{2}\left( \left\vert \mu
\right\vert B^{\prime }\right) ^{5/2}e^{-\pi m^{2}/\left\vert \mu
\right\vert B^{\prime }}I_{b^{\prime }}\,,\ \ I_{b^{\prime
}}=\int_{0}^{\infty }\frac{du}{\left( u+1\right) ^{5/2}}\ln \left( \frac{%
\sqrt{1+u}+\sqrt{1+2u}}{\sqrt{u}}\right) e^{-\pi b^{\prime }u}\,,
\label{3.39b}
\end{equation}%
where$\ b^{\prime }=m^{2}/\left\vert \mu \right\vert B^{\prime }$. For the
sake of comparison with the total number of neutral Fermions created from
the vacuum by a linearly-growing magnetic step \cite{GavGit13}, let us study
the behavior of (\ref{3.39b}) in strong-inhomogeneity and weak-inhomogeneity
cases, specified by the conditions $b^{\prime }\ll 1$ and $b^{\prime }\gg 1$%
, respectively. In the strong-inhomogeneity case, we may expand the
exponential in $I_{b^{\prime }}$ and retain the first terms of the series
to realize that the flux density of the created particles (\ref{3.39b})
admits the form%
\begin{equation}
n^{\mathrm{cr}}=\varrho ^{2}\left( \frac{\pi +\ln 2-1}{6\pi ^{3}}\right)
\left( \left\vert \mu \right\vert B^{\prime }\right) ^{5/2}e^{-\pi
m^{2}/\left\vert \mu \right\vert B^{\prime }}\left[ 1+O\left( \frac{\pi m^{2}%
}{\left\vert \mu \right\vert B^{\prime }}\right) \right] \,,\ \ \frac{m^{2}}{%
\left\vert \mu \right\vert B^{\prime }}\ll 1\,.  \label{3.29.1}
\end{equation}%
This result can be immediately compared to the flux density of the created
particles from the vacuum by a linearly-growing magnetic step in the
strong-inhomogeneity case, found previously in Ref. \cite%
{GavGit13}; cf. Eq. (60). To establish an effective way of comparing results
obtained by different external fields, we consider that both external fields
have the same step magnitude $\mathbb{U}$\ and the same degree of
inhomogeneity, determined by the ratio $m^{2}/\left\vert \mu \right\vert
B^{\prime }$. Rephasing Eq. (\ref{3.29.1}) in terms of $\mathbb{U}$ and $%
m^{2}/\left\vert \mu \right\vert B^{\prime }$,%
\begin{equation}
n^{\mathrm{cr}}=m\mathbb{U}^{2}\left( \frac{\pi +\ln 2-1}{24\pi ^{3}}\right) 
\sqrt{\frac{\left\vert \mu \right\vert B^{\prime }}{m^{2}}}e^{-\pi
m^{2}/\left\vert \mu \right\vert B^{\prime }}\left[ 1+O\left( \frac{\pi m^{2}%
}{\left\vert \mu \right\vert B^{\prime }}\right) \right] \,,\ \ \frac{m^{2}}{%
\left\vert \mu \right\vert B^{\prime }}\ll 1\,,  \label{3.29.2}
\end{equation}%
and comparing with the corresponding result found in Ref. \cite{GavGit13} for a
linearly-growing magnetic step in the same regime\footnote{%
Eq. (\ref{3.29.3}) follows from Eq. (60) in \cite{GavGit13} after summing
over all spin polarizations $s=\pm 1$ and setting $\mathbb{U}=\left\vert \mu
\right\vert B^{\prime }L_{y}$.}%
\begin{equation}
n^{\mathrm{cr}}=m\mathbb{U}^{2}\left( \frac{\sqrt{2}-1+\ln \left( 1+\sqrt{2}%
\right) }{8\pi ^{3}}\right) \sqrt{\frac{\left\vert \mu \right\vert B^{\prime
}}{m^{2}}}e^{-\pi m^{2}/\left\vert \mu \right\vert B^{\prime }}\,,
\label{3.29.3}
\end{equation}%
it is possible to conclude that a Sauter-like magnetic step (\ref{2})
produces fewer pairs from the vacuum compared to a linearly-growing magnetic
step because the numerical factor found in (\ref{3.29.2}), $\approx
3.8\times 10^{-3}$, is smaller than the one appearing in (\ref{3.29.3}), $%
\approx 5.2\times 10^{-3}$. As stated before, this comparison is meaningful
as long as both external fields have the same step magnitude $\mathbb{U}$,
inhomogeneity scale $m^{2}/\left\vert \mu \right\vert B^{\prime }$ and
\textquotedblleft gradually\textquotedblright\ vary along the inhomogeneity
direction. In the case of weak-inhomogeneity, $b^{\prime }=m^{2}/\left\vert
\mu \right\vert B^{\prime }\gg 1$, we may integrate the Laplace-type
integral (\ref{3.39b}) using asymptotic methods \cite{Olver97} to realize
that the flux density of neutral Fermion pairs created is exponentially small%
\begin{eqnarray}
&&n^{\mathrm{cr}}=\frac{\varrho ^{2}}{4\pi ^{4}}\left( \left\vert \mu
\right\vert B^{\prime }\right) ^{5/2}\left( \frac{\left\vert \mu \right\vert
B^{\prime }}{m^{2}}\right) e^{-\pi m^{2}/\left\vert \mu \right\vert
B^{\prime }}\left[ \ln \left( \frac{\pi m^{2}}{\left\vert \mu \right\vert
B^{\prime }}\right) +\ln 4-\psi \left( 1\right) +O\left( \left( \frac{%
\left\vert \mu \right\vert B^{\prime }}{\pi m^{2}}\right) ^{2}\right) \right]
\notag \\
&&\ =\frac{m\mathbb{U}^{2}}{16\pi ^{4}}\left( \frac{\left\vert \mu
\right\vert B^{\prime }}{m^{2}}\right) ^{3/2}e^{-\pi m^{2}/\left\vert \mu
\right\vert B^{\prime }}\left[ \ln \left( \frac{\pi m^{2}}{\left\vert \mu
\right\vert B^{\prime }}\right) +\ln 4-\psi \left( 1\right) +O\left( \left( 
\frac{\left\vert \mu \right\vert B^{\prime }}{\pi m^{2}}\right) ^{2}\right) %
\right] \,,\ \ \frac{m^{2}}{\left\vert \mu \right\vert B^{\prime }}\gg 1\,,
\label{3.29.4}
\end{eqnarray}%
where $\psi \left( 1\right) =-\gamma \approx -0.577$ is Euler's constant and 
$\psi \left( s\right) $ is the Psi (or DiGamma) function \cite{DLMF},%
\begin{equation*}
\psi \left( s\right) =\frac{\Gamma ^{\prime }\left( s\right) }{\Gamma \left(
s\right) }=\frac{1}{\Gamma \left( s\right) }\int_{0}^{\infty
}dxx^{s-1}e^{-x}\ln x\,.
\end{equation*}

At last, one may use the identity $\ln \left( 1-N_{n}^{\mathrm{cr}}\right)
=-\sum_{l=1}^{\infty }\left( N_{n}^{\mathrm{cr}}\right) ^{l}/l$ and perform
integrations similar to the ones discussed before to discover that the
vacuum-vacuum transition probability admits the final form%
\begin{equation}
P_{v}=\exp \left( -\beta V_{y}Tn^{\mathrm{cr}}\right) \,,\ \ \beta
=\sum_{l=0}^{\infty }\frac{\epsilon _{l+1}}{\left( l+1\right) ^{3/2}}\exp
\left( -\frac{l\pi m^{2}}{\left\vert \mu \right\vert B^{\prime }}\right)
\,,\ \ \epsilon _{l}=\frac{I_{b^{\prime }l}}{I_{b^{\prime }}}\,,
\label{3.31}
\end{equation}%
with $n^{\mathrm{cr}}$ given by Eq. (\ref{3.39b}).

It is noteworthy mentioning that relation (\ref{3.22})--which is well-known for strong-field QED with external
electromagnetic fields--holds for the case under consideration as well.
However, a direct similarity of total quantities for both cases is absent.
We see that the flux density of created neutral Fermion pairs and the
quantity $\ln P_{v}^{-1}$\ are quadratic in the magnitude of the step while
the flux density of created charged-particle by the electric step is linear.
This is a consequence of the fact that the number of states with all
possible $\omega $\ and $p_{z}$\ excited by the magnetic-field inhomogeneity
is quadratic in the increment of the kinetic momentum. This is also the
reason why the flux density of created pairs and $\ln P_{v}^{-1}$\ per unit
of the length are not uniform. If the total numbers $V_{y}Tn^{\mathrm{cr}}$\
, given by Eqs. (\ref{3.39b}) and (\ref{3.29.3}), are small, one can use
approximation (\ref{3.39}). This means that the Schwinger effective action
approach works for the case under consideration after a suitable
parameterization.

\subsection{\textquotedblleft \textit{Sharply}\textquotedblright -varying
field configuration\label{Sec3.1.2}}

We now turn the attention to the second configuration of interest, when the
field (\ref{2}) \textquotedblleft sharply\textquotedblright\ steeps near the
origin. Such a configuration is specified by the conditions:%
\begin{equation}
1\gg \sqrt{\varrho \mathbb{U}/2}\gtrsim \frac{m}{\sqrt{\left\vert \mu
\right\vert B^{\prime }}}\,.  \label{3.32}
\end{equation}%
The first inequality indicates that the gradient $\partial _{y}B_{z}\left(
y\right) $ sharply peaks about the origin, while the second implicates that
the Klein zone is relatively small. This configuration is particularly
important due to a close analogy to charged pair production by the Klein
step, see Ref. \cite{DomCal99} for the review. For electric fields whose
spatial inhomogeneity meets conditions equivalent to (\ref{3.32}), it was
demonstrated that the imaginary part of the QED effective action features
properties similar to those of continuous phase transitions \cite%
{GieTor16,GieTor17}. Recently \cite{AdoGavGit20}, we have demonstrated for
the inverse-square electric field that this peculiarity also follows from
the behavior of total quantities when the Klein zone is relatively small.
Because of the condition (\ref{3.32}), not only the parameter $\varrho 
\mathbb{U}/2$ is small but all parameters involving the quantum numbers $%
p_{x}$, $p_{z}$, and $\omega $ are small as well on account of the
inequalities (\ref{3.11}). As a result, the arguments of the hyperbolic
functions in (\ref{3.24}) are small, which means that we may expand the
hyperbolic functions in ascending powers and truncate the corresponding
series to first-order to demonstrate that the mean numbers admit the
approximate form:%
\begin{equation}
N_{n}^{\mathrm{cr}}\approx \frac{4\left\vert p^{\mathrm{R}}\right\vert
\left\vert p^{\mathrm{L}}\right\vert }{\mathbb{U}^{2}-\left( \left\vert p^{%
\mathrm{L}}\right\vert -\left\vert p^{\mathrm{R}}\right\vert \right) ^{2}}\,.
\label{3.33}
\end{equation}%
It is exactly the form of the Klein effect \cite{DomCal99}. Note that unlike
the case of ``gradually"-varying field configuration, exponentially
suppressing factors are absent in (\ref{3.33}). Thus, nontrivial fluxes of
neutral fermions created by the ``sharply''-varying magnetic step can also
be observed. This justifies the study of total quantities when the field sharply varies along the inhomogeneity direction.

To implement the conditions (\ref{3.32}), we conveniently introduce the
Keldysh parameter $\gamma =2m/\mathbb{U}$ and observe that it obeys the
condition $1-\gamma ^{2}\ll 1$ on account of (\ref{3.32}). Next, we perform
the change of variables%
\begin{equation}
\frac{\omega }{m}=\frac{1}{2}\left( 1-\gamma ^{2}\right) \left( 1-v\right)
\,,\ \ \frac{p_{x}^{2}}{m^{2}}=\left( 1-\gamma ^{2}\right) r\,,  \label{3.34}
\end{equation}%
and expand the asymptotic momenta $\left\vert p^{\mathrm{L/R}}\right\vert $
in ascending powers of $1-\gamma ^{2}$ to learn that $\left\vert p^{\mathrm{R%
}}\right\vert /m=\left( 1-\gamma ^{2}\right) ^{1/2}\sqrt{v-r}+O\left( \left(
1-\gamma ^{2}\right) ^{3/2}\right) $, $\left\vert p^{\mathrm{L}}\right\vert
/m=\left( 1-\gamma ^{2}\right) ^{1/2}\sqrt{2-v-r}+O\left( \left( 1-\gamma
^{2}\right) ^{3/2}\right) $. Substituting these approximations into (\ref%
{3.33}) we obtain%
\begin{equation}
N_{n}^{\mathrm{cr}}=\left( 1-\gamma ^{2}\right) \sqrt{\left( 1-r\right)
^{2}-\left( 1-v\right) ^{2}}+O\left( \left( 1-\gamma ^{2}\right) ^{2}\right)
\,.  \label{3.35}
\end{equation}

We now wish to estimate the total number of pairs created from the vacuum by
a sharply varying external field. In this case, it is convenient to first
integrate over $p_{z}$, which is allowed as long as we swap the integration
limits indicated in (\ref{3.29}), i.e. $p_{z}^{\max }=\sqrt{\left( \mathbb{U}%
/2\right) ^{2}-m^{2}-p_{x}^{2}}$ and$\ p_{x}^{\max }=\sqrt{\left( \mathbb{U}%
/2\right) ^{2}-m^{2}}$. Calculating the integral and performing the change
of variables proposed in (\ref{3.34}), we expand the result in power series
of $1-\gamma ^{2}$ to find%
\begin{equation}
\int_{0}^{p_{z}^{\max }}\frac{dp_{z}}{\sqrt{\omega ^{2}+p_{z}^{2}}}=-\frac{1%
}{2}\ln \left( 1-\gamma ^{2}\right) +2\ln 2+\ln \sqrt{1-r}-\ln \sqrt{\left(
1-v\right) ^{2}}+O\left( 1-\gamma ^{2}\right) \,.  \label{3.36}
\end{equation}%
The most significant contribution to total quantities in this regime comes
from the logarithm $\ln \left( 1-\gamma ^{2}\right) $, as $1-\gamma ^{2}\ll
1 $. Neglecting higher-order terms in $1-\gamma ^{2}$, the flux density of
the particles created is approximately given by%
\begin{equation}
n^{\mathrm{cr}}\approx \left( 1-\gamma ^{2}\right) ^{7/2}\left\vert \ln
\left( 1-\gamma ^{2}\right) \right\vert \frac{m^{3}}{\left( 2\pi \right) ^{3}%
}\int_{0}^{r_{\max }}\frac{dr}{\sqrt{r}}\int_{v_{\min }}^{v_{\max }}dv\left(
1-v\right) \sqrt{\left( 1-r\right) ^{2}-\left( 1-v\right) ^{2}}\,,
\label{3.37}
\end{equation}%
where $v_{\min }\approx r$ and $v_{\max }\approx r_{\max }\approx 1$. After
straightforward integrations, the flux density of the particles created from
the vacuum by a sharply varying Sauter-like magnetic step takes the
approximate form%
\begin{equation}
n^{\mathrm{cr}}\approx \frac{4}{105\pi ^{3}}m^{3}\left( 1-\gamma ^{2}\right)
^{7/2}\left\vert \ln \left( 1-\gamma ^{2}\right) \right\vert \,.
\label{3.38}
\end{equation}

Finally, it is important to point out that the behavior of flux densities,
concerning their scaling with parameter $\left( 1-\gamma ^{2}\right) ^{7/2}$%
, can be extended to other types of magnetic steps due to universal forms
for differential quantities in situations where the Klein zone is small.
More precisely, it was found a few years ago \cite{GieTor16,GieTor17} that
the imaginary part of the effective action of QED (both scalar as spinor)
scales with $1-\gamma ^{2}$ in a universal way, irrespective of the
asymptotic behavior of the electric field. Recently \cite{AdoGavGit20}, we
arrived at the same conclusion studying the problem for an specific electric
field and discovered that this compatibility results from the universal
behavior of mean numbers when the Klein zone is small. Inspired by the close
analogy with pure QED and according to peculiarities of differential mean
numbers for sharply varying magnetic steps, we have reasons to believe the
differential mean number of pairs created from the vacuum behaves
universally as Eq. (\ref{3.35}). For these reasons, we suggest that the
imaginary part of the effective action exhibits the universal form given by
Eqs. (\ref{3.39}), (\ref{3.38}), provided the field \textquotedblleft
sharply\textquotedblright\ varies along the inhomogeneity direction.

\subsection{Numerical estimates to the critical field\label{Sec3.2}}

The mechanism here described raises the question about the critical magnetic
field intensity, near which the phenomenon could be observed. It is possible
to estimate such a value based on Fermion's mass and its magnetic moment.
Since $\max B_{z}\left( y\right) =B_{z}\left( +\infty \right) =\varrho
B^{\prime }\equiv B_{\max }$,\ the nontriviality of the Klein zone (\ref%
{3.11}) yields the following condition%
\begin{equation}
\mathbb{U}=2\left\vert \mu \right\vert \varrho B^{\prime }>2m\Rightarrow
B_{\max }>B_{\mathrm{cr}}\,,\ \ B_{\mathrm{cr}}\equiv \frac{m}{\left\vert
\mu \right\vert }\approx 1.73\times 10^{8}\times \left( \frac{m}{1\,\mathrm{%
eV}}\right) \left( \frac{\mu _{\mathrm{B}}}{\left\vert \mu \right\vert }%
\right) \mathrm{G}\,,  \label{4.1}
\end{equation}%
where $\mu _{\mathrm{B}}=e/2m_{e}\approx 5.8\times 10^{-9}\,\mathrm{eV/G}$
is the Bohr magneton \cite{NIST}. For neutrons, whose mass and magnetic
moment are $m_{N}\approx 939.6\times 10^{6}\,\mathrm{eV}$, $\mu _{N}\approx
-1.042\times 10^{-3}\mu _{\mathrm{B}}$, the critical magnetic field (\ref%
{4.1}) is $B_{\mathrm{cr}}\approx 1.56\times 10^{20}\,\mathrm{G}$. More
optimistic values can be estimated for neutrinos because of their light
masses and small magnetic moments. For example, considering recent
constraints for neutrinos effective magnetic moment $\mu _{\nu }\approx
2.9\times 10^{-11}\mu _{\mathrm{B}}$ \cite{gemma12} and mass $m_{\nu
}\approx 10^{-1}\,\mathrm{eV}$ \cite{GiuStu15}, we find $B_{\mathrm{cr}%
}\approx 5.97\times 10^{17}\,\mathrm{G}$. Evidently, this value changes
considering different values to neutrinos' magnetic moment and mass. Taking,
for instance, the experimental estimate to the tau-neutrino magnetic moment $%
\mu _{\tau }\approx 3.9\times 10^{-7}\mu _{\mathrm{B}}$ \cite{donut01} and
assuming its mass $m_{\nu _{\tau }}\approx 10^{-1}\,\mathrm{eV}$ we obtain a
value to $B_{\mathrm{cr}}$ near QED critical field $B_{\mathrm{QED}%
}=m^{2}/e\approx 4.4\times 10^{13}\,\mathrm{G}$, namely $B_{\mathrm{cr}%
}\approx 4.44\times 10^{13}\,\mathrm{G}$. On the other hand, assuming the
lower bound found in Ref. \cite{Bell05} $\mu _{\nu }\approx 10^{-14}\mu _{%
\mathrm{B}}$ and the same mass $m_{\nu }\approx 10^{-1}\,\mathrm{eV}$ we
obtain a value to $B_{\mathrm{cr}}$ orders of magnitude larger than $B_{%
\mathrm{QED}}$, $B_{\mathrm{cr}}\approx 1.73\times 10^{21}\,\mathrm{G}$. The
critical magnetic field surprisingly increases if one considers the magnetic
moment predicted by the SM, $\mu _{\nu }\approx 3.2\times 10^{-19}\mu _{%
\mathrm{B}}\times \left( m_{\nu }/1\,\mathrm{eV}\right) $  \cite{GiuStu09,GiuStu15}. Substituting this
value into (\ref{4.1}) and considering $m_{\nu}\approx 1\,\mathrm{eV}$ we find $B_{\mathrm{cr}%
}\approx 5.41\times 10^{26}\,\mathrm{G}$.

Sterile neutrinos with masses $M$ of several keV are dark matter candidates 
\cite{Kuz09,Drewes17}. Taking into account weak observational constraints on
their magnetic moment $\mu$ \cite{Sig-etal04,Gard09}, one can see that
pairs of sterile neutrinos and antineutrinos could be produced from their
coupling to an inhomogeneous magnetic field. For example, if $M=m_{e}/100$, where $m_e\approx 0.5\,\mathrm{MeV}$ is the electron mass \cite{NIST}, then $\left\vert \mu \right\vert \lesssim 10^{-4}\mu _{B}$ due to precision
electroweak measurements; see e.g. \cite{Gard09}. Hence, we find an estimate to $B_{\mathrm{cr}}$ that is relevant for dark matter, $B_{\mathrm{cr}}\sim 10^{16}\,\mathrm{G}$. These constraints can be weakened
by the mechanism of compositeness and a variety of astrophysical constraints
can be significantly weakened by the candidate particle's mass. For example,
the direct limits on $\left\vert \mu \right\vert $, which would follow from
the nonobservance of Faraday rotation at a given sensitivity, could be $%
\left\vert \mu \right\vert \lesssim \mu _{B}$ \cite{Gard09}. Such a weak
limit implies $B_{\mathrm{cr}}\sim 10^{12}\, \mathrm{G}$.

Besides strong field amplitudes, neutral Fermion pair production requires
inhomogeneous magnetic fields over a certain space area. As discussed in
Sec. \ref{Sec3.1}, the optimal scenario for pair production corresponds to
the case when the field evolves \textquotedblleft
gradually\textquotedblright\ over space. For this configuration, we may
provide a reference value for the field inhomogeneity intensity $B^{\prime }$
based on the properties discussed above. Such a value can be extracted from
condition $\left\vert \mu \right\vert B^{\prime }>m^{2}$, as it supplies an
increase of neutral Fermion pair production according to the analytical
estimate (\ref{3.39b}), for instance. From this condition, we may set the
following reference value $B_{\mathrm{ref}}^{\prime }$%
\begin{equation}
\left\vert \mu \right\vert B^{\prime }>m^{2}\Rightarrow B^{\prime }>B_{%
\mathrm{ref}}^{\prime }\,,\ \ B_{\mathrm{ref}}^{\prime }=\frac{m^{2}}{%
\left\vert \mu \right\vert }\approx 8.77\times 10^{14}\times \left( \frac{m}{%
1\,\mathrm{eV}}\right) ^{2}\left( \frac{\mu _{\mathrm{B}}}{\left\vert \mu
\right\vert }\right) \frac{\mathrm{G}}{\mathrm{m}}\,.  \label{4.2}
\end{equation}%
Considering active neutrinos with mass $m_{\nu }\approx 10^{-1}\,\mathrm{eV}$
and magnetic moment $\mu _{\nu }\approx 2.9\times 10^{-11}\mu _{\mathrm{B}}$%
, we obtain $B_{\mathrm{ref}}^{\prime }\approx 3\times 10^{23}\,\mathrm{G}/%
\mathrm{m}$. This estimate can be decreased by about five orders of
magnitude assuming smaller values to neutrino mass and larger values to its
magnetic moment, say $m_{\nu }\approx 10^{-3}\,\mathrm{eV}$ and $\mu _{\nu
}\approx 1.1\times 10^{-9}%
\mu
_{\mathrm{B}}$ \cite{Lampf}. In the case of sterile neutrinos with masses $M$%
\ of several keV, we obtain $B_{\mathrm{ref}}^{\prime }\sim 10^{21}\,\mathrm{%
G}/\mathrm{m}$.

Lastly, it should be noted that it is also possible to derive an upper bound
to neutrino mass from the same condition given above by assuming a fixed
value to the inhomogeneity size $\varrho $. It was argued in \cite{GiuStu15}
that active neutrino masses should be $m_{\nu }\lesssim 10^{-6}\,\mathrm{eV}$
in order to be created in astrophysical environments filled with magnetic
fields of magnitude $10^{18}\,\mathrm{G}$ and whose inhomogeneity stretches
about one kilometer. As pointed out in \cite{GiuStu15}, this estimate
suggests a profound consideration of theories beyond the SM.

\section{Vacuum fluxes produced \label{Sec4}}

Procedures of renormalization and volume regularization recently presented in Ref. \cite{GavGit21} for strong-field QED with a step potential allows one to calculate and distinguish physical parts of matrix elements of physical quantities given by field operators. Using the technique to map the problem under consideration onto the problem of charged-particle creation by an electric step, we can apply these procedures to find vacuum fluxes of
energy and magnetic moment produced by a magnetic-field inhomogeneity.

One can see from Eq. (\ref{i8}) that $t^{\left( \mathrm{L}\right) }$\ and $%
t^{\left( \mathrm{R}\right) }$\ are macroscopic times of motion of particles
and antiparticles in the remote areas on the left and on the right of the
inhomogeneity, respectively and they are equal,%
\begin{equation}
t^{\left( \mathrm{L}\right) }=t^{\left( \mathrm{R}\right) }=t_{\mathrm{mot}}.
\label{m4}
\end{equation}%
It allows one to introduce an unique time of motion $t_{\mathrm{mot}}$\ for
all the particles in the system under consideration.\ This time can be
interpreted as a time of observation of the evolution of the system under
consideration. The renormalization procedure \cite{GavGit21} allows one to
link quasilocal quantities with observable physical quantities specifying
the vacuum instability.\ In the general case, the matrix elements of
energy-momentum and magnetic momentum operators\ contain local contributions
due to the vacuum polarization and contributions due to the vacuum
instability caused by the external field for all the time $T$\ of his
action.\ We believe that under the condition that that $t^{\left( \mathrm{L}%
\right) }$\ and $t^{\left( \mathrm{R}\right) }$\ are macroscopical, all
local contributions due to the existence of the magnetic-field inhomogeneity
can be neglected.\ Therefore, it is enough to know the longitudinal fluxes
of energy and magnetic moment through the surfaces\ $y=y_{\mathrm{L}%
}\rightarrow -\infty $\ and\ $y=y_{\mathrm{R}}\rightarrow \infty $\ to
construct the initial and final states, link them, and then calculate characteristics of the vacuum instability.

It is clear that such fluxes of created pairs depend on the parameter of the
volume regularization $t_{\mathrm{mot}}$\ due to the presence of the
normalization factor $\mathcal{M}_{n}^{-1/2}$\ in the field operator
decomposition (\ref{3.18}).\ Thus, we can find their relation to observable
physical quantities and obtain a relation between the parameter $t_{\mathrm{%
mot}}$\ and the whole time $T$.\ Such a relation fixes the\ proposed
renormalization procedure.

We suppose that all the measurements are performed during a macroscopic time 
$T$\ when the magnetic field can be considered as constant. In this case,
for example, we believe that the longitudinal flux of the magnetic moment of
particles created with a given $s$\ is equal to the flux density $n_{s}^{%
\mathrm{cr}}$\ of the particles times the modulus of a magnetic moment $%
\left\vert \mu \right\vert $,%
\begin{equation}
\mathcal{M}_{s}^{\mathrm{cr}}=\left\vert \mu \right\vert n_{s}^{\mathrm{cr}}
\label{4a}
\end{equation}%
under the condition that the times $t_{\mathrm{mot}}$\ and $T$\ coincide,
i.e., $t_{\mathrm{mot}}=T$. Such a relation fixes the\ proposed
renormalization procedure.

For the the longitudinal energy flux of particles created with a given $s$\
we find%
\begin{eqnarray}
T_{\mathrm{cr}}^{20}\left( \mathrm{L}\right)  &=&-\sum_{n\in \Omega
_{3}}\left\vert \pi _{s}\left( \mathrm{L}\right) \right\vert N_{n}^{\mathrm{%
cr}},\ \ y=y_{\mathrm{L}}\rightarrow -\infty ,  \notag \\
T_{\mathrm{cr}}^{20}\left( \mathrm{R}\right)  &=&\sum_{n\in \Omega
_{3}}\left\vert \pi _{s}\left( \mathrm{R}\right) \right\vert N_{n}^{\mathrm{%
cr}}\ ,\ \ y=y_{\mathrm{R}}\rightarrow \infty .  \label{4b}
\end{eqnarray}%
The flux density of the particles created with a given $s$\ are equal $%
n_{+1}^{\mathrm{cr}}=n_{-1}^{\mathrm{cr}}$. However, the composition of
these fluxes are different. The magnetic-field inhomogeneity, $\partial _{y}B\left( y\right)
\geq 0$, accelerates particles with $s=-1$\ and antiparticles with $s=+1$\
along the $y$\ axis. At the same time, antiparticles with $s=-1$\ and particles
with $s=+1$\ are accelerated by the field in the opposite direction. Thus, we have
particles with $s=-1$ and antiparticles with $s=+1$ at $y=y_{\mathrm{R}%
}\rightarrow \infty $ while antiparticles with $s=-1$ and particles with $%
s=+1$ appear at $y=y_{\mathrm{L}}\rightarrow -\infty $. The vacuum flux
aimed in one of the directions is formed from fluxes of particles and
antiparticles of equal intensity and with the same magnetic moments parallel
to the magnetic field. In such a flux, particle and antiparticle velocities
that are perpendicular to the plane of the magnetic moment and flux
direction are essentially depressed. Backreaction to the vacuum flux leads
to a smoothing of the magnetic-field inhomogeneity. Such mechanism has to be
taken into account in astrophysics and cosmology.

\section{Concluding remarks\label{Sec5}}

In this work, we study a mechanism that explains the creation of neutral
Fermion pairs with anomalous magnetic moments from the vacuum by
inhomogeneous magnetic fields. We show that solutions of the DP equation
with magnetic-field inhomogeneity can be given in terms of states with
well-defined spin polarization. Such states are localizable and can form
wave packets in a given asymptotic region. In this case, the effect is
similar to the Schwinger effect for charged particles in an electric field
and the problem can be technically reduced to the problem of
charged-particle creation by an electric step for which the nonperturbative
formulation of strong-field QED \cite{GavGit16,GavGit21} can be used.

To study the effects of vacuum instability due to magnetic-field inhomogeneities, we consider a magnetic step that allows solving the DP equation and calculating pertinent quantities when the field lies in particular configurations. We find exact solutions of the DP equation with such a field and study nonperturbative characteristics of
neutral Fermion pair production from the vacuum by the step. We also find
vacuum fluxes of energy and magnetic moments.

The vacuum flux aimed in one of the directions is formed by fluxes of
particles and antiparticles of equal intensity and with the same magnetic
moments parallel to the magnetic field. In such fluxes, particle and
antiparticle velocities that are perpendicular to the plane of the magnetic
moment and the flux direction are essentially depressed. This is a typical
property that can be used to observe their behavior in astrophysical
situations. The backreaction to the vacuum fluxes leads to a smoothing of
the magnetic-field inhomogeneity. Such mechanism has to be taken into
account in astrophysics and cosmology. In particular, it may be relevant to studies on dark matter studies.

Our calculations reveal two peculiar features of neutral Fermion pair
production by inhomogeneous magnetic fields compared to charged pair
production by inhomogeneous electric fields. The first one is that the flux
density of created neutral Fermion pairs is quadratic in the magnitude of
the step while the flux density of charged-particles created by an electric
step is linear. This peculiarity derives from the non-cartesian geometry of
the parameter space formed by the quantum numbers and this feature is
inherent to the dynamics of neutral Fermions with anomalous magnetic moments
in inhomogeneous magnetic fields. This also explains why the flux density of
created pairs per unit of the length are not uniform. In particular, it
means that the Schwinger method of the effective action works for the case
under consideration only after a suitable parameterization.

The second feature worth discussing is the behavior of total quantities when
the field \textquotedblleft sharply" varies. It is exactly the form of the
Klein effect \cite{DomCal99}. Unlike the case of \textquotedblleft
gradually\textquotedblright -varying field configuration, exponentially
suppressing factors are absent in this case. Thus, nontrivial fluxes of
neutral fermions created by the ``sharply''-varying magnetic step can also
be observed. For example, if one compares the flux density of created
neutral Fermion pairs\emph{\ }with the total number of electron-positron
pairs created from the vacuum by inhomogeneous electric fields (given, for
example, by Eq. (88) with $d=4$ in \cite{AdoGavGit20}), we observe two major
differences: the first is the presence of a logarithmic coefficient $%
\left\vert \ln \left( 1-\gamma ^{2}\right) \right\vert $, that can be traced
back to the integration over $p_{z}$ (\ref{3.36}) and therefore does not
depend on the external field. To our knowledge, this term has no precedents
in QED (although a logarithmic coefficient of this type may appear in scalar
QED). The second, and more important, is the value of the scaling (or
critical) exponent seen in (\ref{3.38}). In contrast to QED in $3+1$
dimensions, in which $N^{\mathrm{cr}}\sim \left( 1-\gamma ^{2}\right) ^{3}$ 
\cite{GieTor16,GieTor17,AdoGavGit20}, the total number of neutral Fermions
pairs created from the vacuum features a larger exponent, $7/2$. Aside from
minor numerical differences, this means that the total number (\ref{3.38})
has an extra term $\sqrt{1-\gamma ^{2}}\left\vert \ln \left( 1-\gamma
^{2}\right) \right\vert $, which is always less than unity in the range of
values to $\gamma $ within the interval $0\leq \gamma <1$. Formally, this
indicates that backreaction effects caused by neutral Fermions produced by
sharply-evolving inhomogeneous magnetic fields may be significantly smaller
than backreaction effects expected to occur for QED under equivalent
conditions.

Last but not least, it is necessary to comment on the role of a constant and
homogeneous magnetic field $B_{0}$ when added to the external field (\ref{2}%
). This corresponds to a \textquotedblleft shift\textquotedblright\ of the
Klein zone (\ref{3.11}) and of the distribution (\ref{3.24}) (with respect
to the $\omega $-axis) to the left or to the right, depending on the sign of 
$B_{0}$. Because the step magnitude $\mathbb{U}$ is invariant, the
integration domains in Eqs. (\ref{3.21}), (\ref{3.22}), (\ref{3.29}) remains
unchanged, meaning that the flux density of created pairs $n^{\mathrm{cr}}$
and the vacuum-vacuum transition probability\ $P_{v}$ do not depend on $%
B_{0} $. This is consistent with the heuristic interpretation that a
constant and homogeneous magnetic field does not produce work on particles
with magnetic dipole moment and, therefore, cannot produce pairs from the
vacuum. Evidently, the situation is different for particles having a
magnetic charge because constant and homogeneous magnetic fields produce
work on such particles and may create pairs from the vacuum as discussed in
many references, e.g. \cite{AffMan82,AffAlvMan82,Kobayashi21,LonVac15}. We
hope that the mechanism discussed in this study might offer important
insights to the understanding of neutral Fermion pair production, both in
astrophysical environments as well as by dense matters, which, despite being
genuinely different, shares similarities with the present case if the matter
is time-independent and inhomogeneous.

\section*{Acknowledgements}

The work of S.P. Gavrilov and D.M. Gitman was supported by Russian Science Foundation, grant no. 19-12-00042 (Sections 1,2,4,6,7), and the work of Z.-W. He was supported by the Post-graduate's Innovation Fund Project of Hebei Province, grant No. CXZZSS2021016 (Sections 3,5).

\end{document}